\documentclass[onecolumn, secnumarabic, amssymb, amsmath, superscriptaddress, eqsecnum, nofootinbib]{revtex4-1}

\usepackage{setspace}
\usepackage{braket}
\usepackage{verbatim}
\usepackage{hyperref}
\DeclareMathOperator{\Tr}{Tr}
\DeclareMathOperator{\ex}{ex}
\DeclareMathOperator{\f}{full}
\DeclareMathOperator{\phy}{phy}
\DeclareMathOperator{\can}{can}
\DeclareMathOperator{\cov}{cov}
\DeclareMathOperator{\dep}{dep}
\begin{document}

\title{Statistical Equilibrium in Quantum Gravity: \\
Gibbs states in Group Field Theory}

\author{Isha Kotecha}
\email{isha.kotecha@aei.mpg.de}
\affiliation{Max Planck Institute for Gravitational Physics (Albert Einstein Institute), Am M\"{u}hlenberg 1, 14476 Potsdam-Golm, Germany.}
\affiliation{Institute for Physics, Humboldt-Universit\"{a}t zu Berlin, Newtonstra{\ss}e 15, 12489 Berlin, Germany.}
\author{Daniele Oriti}
\email{daniele.oriti@aei.mpg.de}
\affiliation{Max Planck Institute for Gravitational Physics (Albert Einstein Institute), Am M\"{u}hlenberg 1, 14476 Potsdam-Golm, Germany.}
\affiliation{II Institute for Theoretical Physics, University of Hamburg,  Luruper Chaussee 149, 22761 Hamburg, Germany.}

\begin{abstract}
\vspace{1cm}
Gibbs states are known to play a crucial role in the statistical description of a system with a large number of degrees of freedom. They are expected to be vital also in a quantum gravitational system with many underlying fundamental discrete degrees of freedom. However, due to the absence of well-defined concepts of time and energy in background independent settings, formulating statistical equilibrium in such cases is an open issue. This is even more so in a quantum gravity context that is not based on any of the usual spacetime structures, but on non-spatiotemporal degrees of freedom. In this paper, after having clarified general notions of statistical equilibrium, on which two different construction procedures for Gibbs states can be based, we focus on the group field theory formalism for quantum gravity, whose technical features prove advantageous to the task. We use the operator formulation of group field theory to define its statistical mechanical framework, based on which we construct three concrete examples of Gibbs states. The first is a Gibbs state with respect to a geometric volume operator, which is shown to support condensation to a low-spin phase. This state is not based on a pre-defined symmetry of the system and its construction is via Jaynes' entropy maximisation principle. The second are Gibbs states encoding structural equilibrium with respect to internal translations on the GFT base manifold, and defined via the KMS condition. The third are Gibbs states encoding relational equilibrium with respect to a clock Hamiltonian, obtained by deparametrization with respect to coupled scalar matter fields. 
\end{abstract}

\maketitle
\tableofcontents


\section{Introduction}

The question of how a set of quantum degrees of freedom of spacetime, described by some fundamental dynamical theory, gives rise to the macroscopic continuum spacetime of general relativity, is possibly {\it the} crucial open issue in quantum gravity approaches. In quantum gravity formalisms wherein these fundamental quantum degrees of freedom are of a more exotic nature than quantised geometric fields, it is the problem of the ``emergence'' of spacetime from non-spatiotemporal structures. This is the situation in most modern approaches, including those where quantum gravitational microstates can be formulated as quantum many-body states \cite{oriti2017, hu2005}. Asking this question, then,  inevitably leads one to investigate the collective behaviour of these fundamental micro-constituents. This is the realm of statistical mechanics and field theory. Thus from the perspective of emergent spacetime, the role of statistical methods in quantum gravity is crucial. Besides being instrumental to the issue of spacetime emergence, one should also expect that the collective statistical behaviour of quantum gravity degrees of freedom will produce novel, non-perturbative effects, appearing as modifications of general relativistic dynamics, and relevant for effective gravitational physics.   \\

The specific quantum gravity formalism used in this paper is group field theory (GFT) \cite{krajewski, oritibook2, oritibook1}. With this work we begin investigations into the statistical mechanics of the quanta of group field theory, which are fundamental `atoms' containing discrete gravitational information (as well as discretised matter degrees of freedom, depending on the specific model), and in terms of which quantum spacetime is indeed (tentatively) described as a quantum many-body system, albeit of a very exotic nature. \\

One of the foundational concepts in statistical physics is that of equilibrium. Equilibrium configurations are those that are invariant under time evolution (in turn identified, in flat space, with time translations), generated by the Hamiltonian of the system. But how does one define statistical equilibrium when there is no preferred time and Hamiltonian? This is the case in classical constrained systems such as general relativity. This is also the case in quantum gravitational contexts, especially in formalisms that are not based on continuum spacetime structures, like differentiable spacetime manifolds etc. This is the open problem of defining statistical equilibrium in a (non-spatiotemporal) background independent system. Still, since equilibrium states hold a special place in statistical physics, this is where we start, for developing a statistical mechanical formulation of quantum gravity within a group field theory formalism. \\

In this work we aim to construct Gibbs equilibrium states for a GFT system. Whether these states provide a truly comprehensive characterisation of statistical equilibrium in quantum gravity \emph{in general} is a different (and challenging) issue that is not considered here. We also do not analyse here the physical consequences of our results for a description of (quantum) spacetime or gravity. These are certainly important tasks, which for now are left to future studies.  \\

What we investigate is the statistical mechanics of quanta of spacetime themselves as formulated within GFT, and not the statistical mechanics of matter on a background spacetime of fixed geometry (which is well-understood). Further, this goal should be carefully distinguished from the \emph{closely-related} open issue of formulating a framework for generally covariant statistical mechanics, including that of a dynamical gravitational field. For a nice introduction to these tasks, see \cite{rovelli93, *connesrovelli, rovelli12, montesinosrovelli, chircohaggard} and references therein, which form also the conceptual basis for our work. Specifically, the conceptual issues (such as timelessness) that one faces when investigating the statistical mechanics of general relativistic spacetime, and the statistical mechanics of pre-geometric quanta underlying a spacetime (as defined within a chosen quantum gravity framework) are similar. But formally they are two separate issues, even if related. We deal with the latter. This comes with additional difficulties in principle due to the lack of familiar spatiotemporal structures. For example, generic configurations of our quantum gravity system do not admit an interpretation as quantised geometric fields. Therefore, geometric configurations cannot be presumed, and one would have to look for such phases within the full statistical description of the quanta of spacetime. \\

That being said the present work could contribute, even if only implicitly, to the understanding of the problem of defining a generally covariant statistical mechanics in the gravitational context, thus including the continuum (quantum) gravitational field. But, to spell out this implication and see the consequences for GR and spacetime physics, one would need to solve the issue of emergence from this quantum gravity formalism. This is of course a difficult and an open problem, and also one that we do not directly tackle in this paper, even though our results may contribute to its solution by providing some useful formal tools. This naturally does not mean that we are solving all related issues, neither about statistical mechanics for the gravitational field or in a background independent context, nor about quantum gravity. In fact, the results of this work should be viewed as implementing suitable definitions of statistical equilibrium, and subsequently offering concrete examples of Gibbs states, in a (non-spatiotemporal) background independent quantum gravity context based on insights from previous works (\cite{rovelli93, *connesrovelli, rovelli12} and related others) in covariant (spatiotemporal) settings; and as commencing investigations into the statistical mechanics of group field theory systems. \\

Overall, the perspective that we hold in order to construct a quantum statistical mechanical framework for GFT, which is used thereafter to construct Gibbs states, is to reformulate the system as a many-body quantum system, where instead of chemical atoms or molecules, we deal with fundamental, pre-geometric `particles' that carry gravitational and matter degrees of freedom. Then, once we establish the groundwork for organising its states in an appropriate Hilbert space along with the relevant algebra of observables, we define statistical states as density operators. In the spin network picture, these density operators define statistical states of a system of arbitrarily large spin networks, including disconnected configurations, with a variable but finite number of nodes of fixed valence. The goal is to take advantage of the technical tools made available by the GFT formulation of quantum gravity (spin network) degrees of freedom to apply close-to-standard definitions of equilibrium to define Gibbs states in a fully background independent context, for the very fundamental (candidate) building blocks of quantum spacetime, and within the full theory (as opposed to special approximations). Specifically, these tools offer advantages at two levels. First, the suggested formal description of spacetime as a many-body quantum system allows us to handle these issues within a mathematical formalism that maintains close analogies with that used for more mundane physical systems. This, in a way, permits us to move forward even without having fully solved all the conceptual issues implicated in the problem. Second, while GFTs are fully background independent from the point of view of spacetime physics (spacetime itself has to be `reconstructed' in most of its features), their mathematical definition as field theories on Lie group manifolds allows us to work with the background structures \emph{of the group manifold} playing technically a very similar role to what spacetime structures (spacetime metric, topology, etc) play in usual field theories, e.g. for condensed matter systems.   \\

The paper is organised as follows. Section \ref{backgroundindependence} discusses the issue of statistical mechanics and background independence, and its relevance to group field theories, specifically to put into context the work of this paper. In section \ref{gft} the construction of group field theories is presented. With the relevant structures and definitions in place, in section \ref{gfteqm} we present a quantum statistical mechanical framework for the Fock representation of GFTs and subsequently construct examples of model-independent, structural Gibbs states in sections \ref{sectionvolgibbs} and \ref{kmsgibbs}. In section \ref{physeqm} we give a third definition of a Gibbs state of relational type, based on deparametrization of the original GFT system to define a canonical system with a clock structure. Finally, we summarise our results and offer some outlook.


\section{Background independence, statistical equilibrium and Gibbs states} \label{backgroundindependence}

What characterises statistical equilibrium? In a non-relativistic system, the answer is unambiguous. Equilibrium states are those which are stable under time evolution generated by the Hamiltonian $\hat{H}$ of the system. In the algebraic description, this property is possessed by states which satisfy the Kubo-Martin-Schwinger (KMS) condition \cite{kubo, *martinschwinger, HHW1967}. For finite systems, KMS states take the explicit form of Gibbs states, whose density operators have the standard form proportional to $e^{-\beta\hat{H}}$. This characterisation of equilibrium is unambiguous because of the special role played by time and its conjugate energy in non-relativistic mechanics, where time is absolute, modelled as the unique, external parameter encoding the dynamics of the system.  \\

Investigating this question in a background independent context becomes much more challenging and interesting, and a complete framework for statistical mechanics in this setting is still missing. The primary reason is the modified role that time plays in such theories \cite{rovellitime1, *rovellitime2}. Classical gravity as described by GR is diffeomorphism invariant (aka generally covariant). This means that space and time coordinates have no physical significance. They are simply gauge. More physically, all geometric quantities, in particular temporal intervals, are dynamical, and generic solutions of the GR dynamics do not allow to single out any preferred time (or space, for what matters) direction. This is the content of background independence in GR, and other modified gravity theories with the same symmetry content.  Specifically, the time coordinate is no longer a universal, physical evolution parameter. In quantum gravity formalisms in which an even more radical setup is invoked, in which even the familiar spatiotemporal structures of GR like the differential manifold, continuum metric and matter fields, etc have disappeared, the absence of an unambiguous notion of time evolution is even more conspicuous.  How can one define a thermal (statistical) state and specifically, an equilibrium state, then? \\

There are different proposals in the literature for a definition of statistical equilibrium that could be general enough to apply to generally covariant systems; and (independent of the exact context, be it classical or quantum, particle mechanics or field theory) these can be observed to be based on different conceptual underpinnings (and often on a combination of the same) that characterise the well-understood non-relativistic equilibrium configurations. The following are few of these principles: KMS condition and Tomita-Takesaki theory (`thermal time' hypothesis) \cite{rovelli93, connesrovelli, rovelli12}, where in some sense the problem is turned upside down, with a suitable identification of an equilibrium state used to {\it define} a notion of time, adapted to that state; ergodic principle \cite{chircojosset, montesinosrovelli}; principle of optimisation of a relevant thermodynamic potential (entropy or free energy) \cite{montesinosrovelli}; nature of the split into and interactions among the subsystems responsible for thermalisation \cite{chircojosset, haggardrovelli, chircohaggard, chircojosset16}.  \\

Group field theories are also background independent in the radical sense specified above for `spacetime-free' quantum gravity formalisms, but they also present specific peculiarities, which will be crucial in our analysis. The base space for the GFT fields consists of Lie group manifolds, encoding discrete geometric as well as matter degrees of freedom. This is \emph{not} spacetime, and all the usual spatiotemporal features associated with the base space of a standard field theory are absent. As in other covariant formalisms, a physically sensible strategy is to use internal dynamical variables, for example matter fields, as relational clocks with respect to which one defines evolution. Even in this case though, one should not expect the existence of a preferred material clock, nor, having chosen one, that this would provide a  `perfect' clock, mimicking precisely an absolute time coordinate. In the end, like standard constrained systems, GFTs too are devoid of an external or even an internal variable that is clearly identified as a preferred evolution parameter. However, the close-to-standard QFT language used in GFTs, with its Fock space and in particular the presence of a base manifold (the Lie group, with associated metric and topology) imply the availability of some mathematical structures that are crucially shared with spacetime-based QFTs; this is a main advantage over other QG formalisms.
\\ 
 

In this paper, we consider Gibbs states as the relevant equilibrium states, i.e. states of the exponential form $e^{-{\mathcal{O}}}$ (for some ${\mathcal{O}}$ not necessarily a Hamiltonian). Before delving into the details of the GFT formalism and how we define such states within it, we discuss below the general ideas guiding our construction. The following discussion is not restricted to the group field theory formalism, or even to classical or quantum sectors. Rather it attempts to present in a coherent way the perspectives and strategies employed in past studies (\cite{rovelli93, connesrovelli, rovelli12, chircojosset, montesinosrovelli, haggardrovelli, chircohaggard} and related works) for defining Gibbs states and statistical equilibrium in a background independent system.  \\

Gibbs statistical states can be categorised according to two main criteria: \\
A. whether the state is a result of considering an associated pre-defined flow/transformation of the system or not; \\
B. the nature of the functions or operators (in the exponent) characterising the state, specifically whether these quantities encode the physical dynamics of the system, thus being in this sense model-dependent, or not. \\

Categories A and B are mutually independent, in the sense that a single system could simultaneously be both of types A and B. Details of these two categories, their respective subclasses and related examples follow. \\

Let us first look at A, which can be considered at a slightly higher footing than B because the contents of classification under this category are the actual construction procedures or the `recipes' used to arrive at a resultant Gibbs state. Moreover, it is within A where we observe that the well-known Jaynes' entropy maximisation principle \cite{jaynes} could prove to be especially useful in background independent contexts such as in non-perturbative quantum gravity frameworks. Under A, we can identify two recipes or ways with which to characterise a Gibbs state depending on the information at hand for a given system. \\

\textbf{A1. Dynamical: Use of KMS condition} \\

The KMS condition \cite{kubo, *martinschwinger, HHW1967, robinson} is formulated in terms of a 1-parameter group of automorphisms of the system. A KMS state encodes stationarity with respect to this 1-parameter flow. Thus, if in the given description of a system, one can identify a relevant set of transformations with respect to which one is interested in defining an equilibrium state, then one asks for the state to satisfy the KMS condition with respect to a 1-parameter (sub-)group of the said transformations to arrive (for finite systems) at a Gibbs state $\rho \propto e^{-\beta \mathcal{G}}$, where $\mathcal{G}$ is the generator of the flow. The (inverse) `temperature' $\beta$ enters formally as the periodicity in the flow parameter, regardless of the interpretation of the latter. \\

Thus, this characterisation is strictly based on the existence of a suitable pre-defined flow of the configurations of the system and then imposing the KMS condition with respect to it. These transformations could correspond to physical or structural properties of the system (see the discussion of category B below). Simple examples are, respectively, the physical time flow $e^{iHt}$ in a non-relativistic system where $H$ is the Hamiltonian, which gives rise to an equilibrium state $e^{-\beta H}$; and a $U(1)$ gauge flow $e^{i N \theta}$ where $N$ is the number operator, which leads to an equilibrium state $e^{-\beta N}$. \\
 
\textbf{A2. Thermodynamical: Use of Jaynes' constrained entropy maximisation principle} \\

Consider a situation wherein the given description of a system does not include relevant symmetry transformations, or (even if such symmetries exist, which they usually do) that we are interested in those properties of the system which are not naturally associated to sensible flows, in the precise sense of being generators of these flows. An example of the latter is a geometric operator such as area or volume. These are of special interest in the context of quantum gravity since they may be instrumental for statistically extracting macroscopic geometric features of spacetime regions from quantum gravity microstates. In such cases then, what characterises a Gibbs state and what is the notion of `equilibrium' encoded in it? \\

In order to construct a Gibbs state here, where we may only have access to a set of constraints fixing the mean values of a set of functions or operators $\{\langle \mathcal{O}_l \rangle_\rho = \overline{\mathcal{O}_l}\}_{l=1,2,...}$ (in classical or quantum descriptions respectively), one must rely on Jaynes' principle \cite{jaynes} of maximising the entropy $S[\rho] = -\langle \ln \rho \rangle_\rho$ while simultaneously satisfying the above constraints, via the method of Lagrange multipliers. As is standard, the angular brackets here denote the statistical average in a statistical state $\rho$ defined on the state space (be it a phase space in the classical description or a Hilbert space in the quantum description) of the system. Undertaking this procedure, one arrives at a Gibbs state $\rho = e^{-\sum_l \beta_l \mathcal{O}_l}$ (where one of the $\mathcal{O}$'s is the identity fixing the normalisation of the state). Here the `temperatures' $\beta_l$ enter formally as Lagrange multipliers. The averages $\overline{\mathcal{O}_l}$ with parameters $\beta_l$, and other quantities derived from them, can be understood as thermodynamic variables defining a macrostate of the system, and can take on the same formal roles as in usual statistical mechanics and thermodynamics. But their exact interpretation would depend on the context. The identification and interpretation of such \emph{relevant} quantities \emph{is} in fact the non-trivial aspect of the problem, particularly in quantum gravity. \\

This characterisation is strictly independent of the existence of any pre-defined transformations or symmetries of the underlying microscopic system, as long as there is at least one function or operator (identified as relevant) whose statistical average is assumed (or known) to be fixed at a certain value. Consequently, this characterisation could be most useful in background independent settings, exactly since it is based purely on information-theoretic methods, in the same spirit as introduced by Jaynes. Finally, we note that in this characterisation a notion of equilibrium is implicit in the requirement that a certain set of observable averages remain constant, i.e. it is implicit in the existence of the constraints $\langle \mathcal{O}_l \rangle_\rho = \overline{\mathcal{O}_l}$ (see for example the discussion in section II in \cite{tribus}). \\

\bigskip

Let us summarise the above classifications and make additional remarks about category A. The aim is to construct Gibbs states for a system of many quanta (whatever they may be), and the two classifications, dynamical and thermodynamical, under category A offer us two formally independent strategies to do so. Based on our knowledge of the system, we may prefer to use one over the other. If there is a known set of symmetries with respect to which one is looking to define equilibrium, then the technical route one takes is to construct a state satisfying the KMS condition with respect to (a 1-parameter subgroup of) the symmetry group. The result of using this recipe (in a finite system) is a Gibbs density operator $e^{-\beta \mathcal{G}}$, characterised by the generator $\mathcal{G}$ of the 1-parameter flow of these symmetry transformations. On the other hand, if one does not have interest in or access to any particular transformations or flows of the system, but has a partial knowledge about the system in terms of a set of observable functions or operators whose statistical averages are fixed to certain values $\langle \mathcal{O}_l \rangle_\rho = \overline{\mathcal{O}_l}$, then one employs Jaynes' principle of maximising the (Shannon or von Neumann) entropy under the given set of constraints. The resultant statistical state is again of a Gibbs exponential form $e^{-\sum_l \beta_l \mathcal{O}_l}$, which is now characterised by the set of observables $\mathcal{O}_l$.  \\ 

It is important to remark on a particular subtlety. Given a Gibbs state, constructed say from recipe A1, then once it is already defined, it also satisfies the thermodynamic condition of maximum entropy. Similarly, a Gibbs state defined on the basis of A2, after it is constructed also satisfies the KMS condition with respect to a flow that is derived from the state itself. Let's look at a couple of examples. Consider the standard non-relativistic Gibbs state $\propto e^{-\beta H}$ which is constructed by satisfying the KMS condition with respect to unitary time translations $e^{iHt}$. That is, this state is classified as A1 since its construction relies on the KMS condition. But, once this state exists, it is also the one that maximises the entropy under the constraint $\langle H \rangle = E$, along with a normalisation condition. Now consider an example of a state $\propto e^{-\beta V}$, where $V$ is say a geometric volume observable. This state is derived as a result of maximising the entropy under the constraint $\langle V \rangle = v$ (and normalisation), hence it is classified as A2. Once this state is defined, one can \emph{extract} a flow \emph{from} the state, with respect to which it will satisfy the KMS condition.\footnote{In the context of general covariant statistical mechanics, the utility of this observation has been presented in \cite{rovelli93, connesrovelli, rovelli12}, and is the crux of the thermal time hypothesis.} This is the modular flow $e^{i\beta V \tau}$, where $\tau$ is the modular flow parameter.\footnote{In a classical phase space description of a system, the modular flow is the integral curve of the vector field $X_V$ defined by $\omega(X_V) = -dV$, where $\omega$ is the symplectic form and $V$ is a smooth function. In this case naturally the flow is not written in terms of a unitary operator (as done above), but as the vector field $\partial_\tau$. In the quantum C* algebraic description (or in a specific Hilbert space representation of it), the modular flow is that of the Tomita-Takesaki theory.} Therefore, classifications A1 and A2 refer to the construction procedures employed as per the situation at hand. Once a Gibbs state is constructed using any one of the two procedures, then technically it will satisfy both the KMS condition with respect to a flow,\footnote{Whether the flow parameter has a reasonable physical interpretation is a separate issue, and would be expected to depend on the specific context.} and maximisation of the thermodynamic entropy. \\

Now, given that a Gibbs state can be constructed as a result of either the dynamical or the thermodynamical recipe, one can consider the nature of the functions or operators $\mathcal{O}$ that characterise it. This is the content of classification under category B. \\

\textbf{B1. Physical}: $\mathcal{O}$ is associated with the physical dynamics of the system under consideration, i.e. it depends on a specific choice of dynamical equations of motion, thus in this sense is model-dependent. In the non-covariant setting, $\mathcal{O}$ would simply be the Hamiltonian of the system. In a covariant setting, it is more subtle. For a covariant system that is deparametrizable with a suitable choice of a good clock (see section \ref{physeqm} for discussion), then $\mathcal{O}$ would be the associated clock Hamiltonian. Overall, this particular classification refers explicitly, and in the definition of the Gibbs state (thus, before extracting physical {\it consequences} of the given definition) to the physical dynamics of the system, as encoded in a relevant model-dependent function or operator, whether it be a conventional Hamiltonian or a constraint. \\

\textbf{B2. Structural}:  $\mathcal{O}$ are model-independent quantities that do not refer directly to the specific physics of the system. Examples of structural transformations are generic rotations or translations of the base manifold of the theory. Examples of structural quantities not directly associated to transformations would be geometric observables like area or volume. \\

Four different types of Gibbs states can be constructed from combinations of classifications under categories A and B. Let us give some examples, including the ones we will construct in the following. \textbf{A1-B1:} Gibbs states with respect to physical time translations, as considered in standard non-relativistic statistical mechanics; Gibbs states in a system deparametrized with respect to a relational clock time, as will be constructed in section \ref{physeqm} in the case of GFT. \textbf{A2-B1:} examples that are considered in \cite{montesinosrovelli}\footnote{It must be noted that in the examples in this reference the initial identification of which quantities $\mathcal{O}$ are relevant is still based on requiring stationarity along the orbits of some gauge symmetries.}; in the context of GFT, this type of state as defined through the use of a dynamical constraint would correspond to a statistical mechanical reformulation of the partition function for spin network states \cite{oritifock}, which is the focus of a parallel work \cite{kotechainprep}. \textbf{A1-B2:} early examples in a non-gravitational setting belong to the work by Souriau (see recent review \cite{marlereview}); in section \ref{kmsgibbs} of the present work we construct Gibbs states (which are shown to be the unique KMS states in our setting) corresponding to translation automorphisms of the full GFT algebra (as introduced in section \ref{weylGFT}). \textbf{A2-B2:} such states are considered in \cite{montesinosrovelli, krasnov96} with regards to geometric area and volume operators; here in section \ref{sectionvolgibbs}, we construct a Gibbs state characterised by a geometric volume operator defined on the GFT Hilbert space.  \\


Note that for real systems \cite{sewellbook} of finite size, the Gibbs state characterisations of A1-B1 and A2-B1 are equivalent. For infinite systems, there are configurations wherein the two are not equivalent. These are called metastable states which are locally at equilibrium (satisfy the KMS condition) but not globally (do not minimise the free energy density of the system). This typically occurs in systems with long-range interactions. Also for infinite systems, it is well-known that equilibrium states are not of the exponential Gibbs form, and are algebraic states that are more generally characterised by the KMS condition (and entropy maximisation for global equilibrium configurations).


\section{Group field theory} \label{gft}

A group field theory \cite{krajewski,  oritibook1, oritibook2} describes a system of dynamical fields, for which the configuration space consists of suitable group manifolds, instead of spacetime as is the case for standard field theories. The dynamics is encoded in an action functional. The simplest class of models are scalar theories, with fields $\varphi: G^{d} \rightarrow \mathbb{C}$ defined on (several copies of) the local gauge group of gravity $G$. This is the Lorentz group $SL(2,\mathbb{C})$ in 4d or its Euclidean counterpart $Spin(4)$. $SU(2)$ is often used as the relevant subgroup in the context of quantum gravity, especially for models connected to loop quantum gravity\footnote{GFT models of quantum gravity are most often defined to possess an additional symmetry, implemented at the level of each GFT field, which justifies the above discrete geometric interpretation. This is the so-called \lq closure constraint\rq, which is implemented as a gauge invariance of each GFT field under the diagonal right (or left) action of the group $G$, effectively reducing the domain manifold to $G^{\times d}/G$. While conceptually and physically important, this property is not crucial for the problem we tackle in this paper, nor for our mathematical construction of equilibrium states, which can be easily adapted to it. Therefore, we do not deal with it explicitly in the following.}. \\

The quanta of the corresponding quantum GFT field are interpreted (under a set of geometricity conditions) to be polyhedra \cite{bianchi2011} with $d$ number of faces. In the most relevant case of quantum gravity in 4 (would-be) spacetime dimensions, the most developed models are based on simplicial structures and the GFT quanta are interpreted as tetrahedra. In the dual pictorial representation, the same quanta are $d$-valent nodes with open links. The $d$ links are labelled by group elements $\underline{g} \equiv (g_1,g_2,...,g_d) \in G^{d}$ which can be understood (again, in quantum gravity models based on simplicial structures) as parallel transports of the gravitational $G$-connection (discretised on the links of a cellular 2-complex). Equivalently, the $d$ faces of the polyhedra can be labelled by the Lie algebra-valued flux variables \cite{baratin2010, guedes2013},  which are the fluxes of the conjugate tetrad field (discretised on the faces of the dual polyhedral complex), conjugate to the group elements.\footnote{Even though these geometric degrees of freedom, group and flux variables, are understood as eventually generating the continuum connection and tetrad fields respectively, we note that their occurrence and definition within group field theory is independent of any embedding into continuum structures via a discretisation. This is one reason why, unlike in LQG, one does not impose on the GFT states any cylindrical consistency conditions, at least in the definition of the theory. Instead, the working philosophy in GFT is that the corresponding continuum structures be generated dynamically.} The fluxes encode the volumes of the faces. Such algebraic data labelling the GFT quanta or `particles' thus encode (discrete) geometrical information. Contrast this with standard field theories where the particles are labelled by spacetime coordinates, identifying points in a manifold of fixed geometry. \\

With these fundamental `atoms', extended space (kinematical states via composition) and spacetime (dynamical processes via interactions) with arbitrary geometry and topology can in principle be constructed.  Several polyhedra can be glued along shared faces to form a polyhedral complex. A relevant example is again in simplicial models for 4d gravity, where tetrahedra combine along shared triangles to form a 3-dimensional simplicial complex. In the language of quantised group field theory, this corresponds to a quantum state made up of several GFT quanta, i.e. a `multi-particle' state achieved via composition of several single-particle (tetrahedron) states (the gluing being encoded in suitable restrictions on the multi-particle wave function). In the dual picture, a multi-particle GFT state corresponds to a collection of nodes with labelled links, which when glued to each other form an extended labelled graph with valence $d$. In the canonical gravity literature, this is simply a spin network of a $G$-connection \cite{rovellibook}. Furthermore, the interactions of the quanta as dictated by the action gives rise to a discretised spacetime or a spinfoam. The picture that we have is thus of a system of quanta whose states and processes as defined by the associated group field theory corresponds to discretised spacetimes of arbitrary geometry and topology. A group field theory is thus a quantum field theory of discrete geometrical building blocks (for simplicial interactions, of simplicial geometry) \cite{oritibook1, oritibook2, oritibook3}. \\

The above description concerns models of \lq pure geometry\rq. In the present work, we are concerned also with group field theories which generate discrete spacetimes coupled to (discretised) real scalar matter fields. One way to couple a single real scalar degree of freedom, used in recent GFT literature, in particular for cosmological applications \cite{oritireview2016,ewing2016,gielen2017}, and detailed in \cite{zhang2017}, is by extending the original configuration space $G^d$, encoding purely geometrical data, by $\mathbb{R}$. By extension, $n$ number of scalar fields can be coupled by considering a model based on $G^d \times \mathbb{R}^n$. Doing this we have assigned to each GFT particle additional $n$ real numbers representing the values of the fields. Consequently, a GFT Feynman diagram ($d$-dim polyhedral complex) is enriched by $n$ scalar fields that are discretised on the vertices of the graph dual to $(d-1)$-dim hypersurfaces. Thus we are concerned with a GFT field defined (for arbitrary natural numbers $d>0$ and $n\geq0$) as,
\begin{equation}
\varphi: G^{d} \times \mathbb{R}^{n} \rightarrow \mathbb{C} \;.
\end{equation}
The primary reason for including matter in GFTs is obvious: any fundamental theory of gravity must include (or must be able to generate at an effective level) matter degrees of freedom if it is to eventually realistically describe the universe. Also, as discussed in the previous section, in background independent systems like GFTs, as is well-known \cite{rovelli1, *smolinrovelli, *brownkuchar, *tamborninoreview}, a reasonable way of defining physical quantities is by constructing relational observables using material reference frames. In GFT, such relational reference frames using scalar fields have indeed been used in the context of cosmology \cite{gielen2017, ewing2016, oritireview2016}. In the present work, we use them for a different, if related, purpose. In section \ref{kmsgibbs} we define Gibbs states with respect to internal translations in these scalar fields within the framework of quantised GFT. Later on in section \ref{physeqm}, we consider the issue of deparametrizing the full GFT system with respect to one of these several scalar fields to define relational dynamics. The resulting relational clock Hamiltonian is used to construct physical relational equilibrium states.


\subsection{Fock space and GFT algebra} \label{fockspace}

Adopting the second quantisation scheme \cite{robinson, fetter}, multi-particle states of the quantum field $\varphi(\underline{g},\underline{\phi})$ can be organised in a Fock space $\mathcal{H}_F$ generated by a Fock vacuum $\ket{\Omega_F}$ and the ladder operators\footnote{For most part of this paper since we work only in the quantum setting, we omit the use of a hat $\;\widehat{}\;$ to distinguish an operator, for convenience. But since Section \ref{physeqm} considers both the classical and quantum sides, there hats are employed to avoid confusion.} $\varphi,\varphi^*$ associated with the GFT field \cite{oritifock, mikovic}. A single quantum is created by acting on the Fock vacuum with the creation operator
\begin{equation}
\varphi^*(\underline{g},\underline{\phi}) \ket{\Omega_F} = \ket{\underline{g},\underline{\phi}} \;. 
\end{equation}
The Fock vacuum is the state with no quantum geometrical or matter degrees of freedom, satisfying $\varphi(\underline{g},\underline{\phi}) \ket{\Omega_F} = 0$ for all arguments. $\ket{\underline{g},\underline{\phi}}$ is the state of a $d$-valent node whose links are labelled by group elements $\underline{g} \equiv (g_1,...,g_d)$ and the node itself with a set of real numbers $\underline{\phi}\equiv (\phi_1,...,\phi_n)$. Then, a generic single-particle state with wavefunction $\psi$ is given by, 
\begin{equation*}
\ket{\psi} = \int_{G^d} d\underline{g} \int_{\mathbb{R}^n} d\underline{\phi} \; \psi(\underline{g},\underline{\phi}) \, \ket{\underline{g},\underline{\phi}},
\end{equation*}
where $d\underline{g} = \prod_{I=1}^d dg_I$ is the Haar measure\footnote{$G$ is taken to be locally compact so that the Haar measure is defined, even though it would be finite only for compact groups. We also require $G$ to be unimodular (for a later proof). Both these properties are satisfied by the physically relevant cases of $SL(2,\mathbb{C}), Spin(4)$ and $SU(2)$.}, $d\underline{\phi} = \prod_{a=1}^n d\phi_a$ is the Lebesgue measure, and $\psi$ is an element of the single-particle Hilbert space\footnote{Even for compact $G$, the configuration space is non-compact along $\mathbb{R}$. Regularisation then requires restricting to a compact domain. We shall return to this when discussing the GFT Weyl algebra in the next section.} $\mathcal{H} = L^2(G^d \times \mathbb{R}^n)$. For $d=4$, this is the space of states of a quantum tetrahedron \cite{barbieri, *baezbarrett} with additional real numbers attached to it. \\ 

The complete pre-Fock space is $\bigoplus_{N \geq 0} \mathcal{H}^{\otimes N}$, where $\mathcal{H}^{\otimes N}$ describes the $N$-particle sector. The Fock space is the symmetric projection of this. We (choose to) impose on our states symmetry under arbitrary particle exchanges, i.e. bosonic statistics. In the spin network picture, this condition reflects the graph automorphism of vertex relabelling and is a natural feature to require. For the case at hand then, the Hilbert space for bosonic GFT quanta is the Fock space
\begin{equation}
\mathcal{H}_F = \bigoplus_{N \geq 0} \text{sym}\, \mathcal{H}^{\otimes N} \;.
\end{equation}

We stress again that this Fock space contains arbitrary spin network excitations \cite{oritifock}, thus all the quantum gravity structures are shared with loop quantum gravity (even if organised in a different way). This means that defining proper statistical equilibrium states on this Fock space truly means defining non-perturbative statistical equilibrium states in a fully background independent context and within a fundamental theory of quantum gravity based on spin network states.  \\

The ladder operators (which take us between the different multi-particle sectors) satisfy the commutation relations (CR) algebra\footnote{These commutation relations would look slightly different if the closure condition is imposed on GFT fields and quanta.},
\begin{equation} \label{ladderCR}
[\varphi(\underline{g}_1,\underline{\phi}_1), \varphi^*(\underline{g}_2, \underline{\phi}_2)] = \mathbb{I}(\underline{g}_1,\underline{g}_2)\delta(\underline{\phi}_1-\underline{\phi}_2) \;\;,\;\; [\varphi, \varphi] = [\varphi^*, \varphi^*] = 0
\end{equation}
where, $\mathbb{I}$ and $\delta$ are delta distributions for functions on $G^d$ and $\mathbb{R}^n$. \\


\textbf{Spin-momentum basis.} The spin-momentum basis, which is a result of harmonic analysis on $G^d \times \mathbb{R}^n$, is used in the examples considered later and is thus summarised here. The GFT field transforms in the familiar way,
\begin{equation}
\varphi(\underline{g},\underline{\phi}) = \int_{\mathbb{R}^n} \frac{d\underline{p}}{(2\pi)^n} \sum_{\underline{\chi} \in \text{Irreps}(G)} D_{\underline{\chi}}(\underline{g}) \; e^{-i\underline{p}.\underline{\phi}} \; \varphi(\underline{\chi},\underline{p}) 
\end{equation} 
where $D_{\underline{\chi}}(\underline{g})$ are normalised Wigner expansion modes. The corresponding ladder operators would also transform in the same way. Formally $p$ enters as the Fourier conjugate to $\phi$ but in light of interpreting $\phi$ as a scalar field, $p$ would be its corresponding observable momentum. Modes $\underline{\chi}$ are the data characterising the irreducible representations of $G$. For example, for $G=SU(2)$, the modes $\underline{\chi} = (\underline{J},\underline{m},\underline{n})$ (or $(\underline{J},\underline{m},\mathcal{I})$ for gauge-invariant $\varphi$, where $\mathcal{I}$ is the intertwiner basis) denote the usual spin data of a single node. \\

The algebra structure \eqref{ladderCR} is preserved, and now takes the form,
\begin{equation} \label{spinCR}
[\varphi(\underline{\chi}_1,\underline{p}_1), \varphi^*(\underline{\chi}_2,\underline{p}_2)] = (2\pi)^n \delta_{\underline{\chi}_1 \underline{\chi}_2} \delta(\underline{p}_1 - \underline{p}_2) \;\;,\;\; [\varphi, \varphi] = [\varphi^*, \varphi^*] =0
\end{equation}
where $\delta_{\underline{\chi}_1 \underline{\chi}_2}$ is the Kronecker delta, $\delta(\underline{p}_1 - \underline{p}_2)$ is the Dirac delta distribution on $\mathbb{R}^n$, and $\varphi(\underline{\chi},\underline{p})\ket{\Omega_F} = 0$ for all $\underline{\chi},\underline{p}$. \\


\textbf{Occupation number basis.} The Fock space basis that is utilised later in the paper is the orthonormal occupation number basis \cite{oritifock, fetter}. It is particularly useful because it is the eigenbasis of the number operator in $\mathcal{H}_F$. This basis sees only how many particles occupy a given mode $\underline{\chi}$. Utilising the CR relations \eqref{spinCR}, a normalised multi-particle state with $n_{\underline{\chi}}$ number of particles in a single mode $\underline{\chi}$ is,
\begin{equation}
\ket{n_{\underline{\chi}}} = \frac{1}{\sqrt{n_{\underline{\chi}}!}} (\varphi^*_{\underline{\chi}})^{n_{\underline{\chi}}}\ket{\Omega_F} \;.
\end{equation}
Then, a generic multi-particle state occupying several modes $\underline{\chi}_i$ is,
\begin{equation}
\ket{\{n_{\underline{\chi}_i}\}} \equiv \ket{n_{\underline{\chi}_1}, n_{\underline{\chi}_2}, ..., n_{\underline{\chi}_i}, ...} = \frac{1}{\sqrt{\prod_i \left(n_{\underline{\chi}_i}!\right)}} \prod_i (\varphi^*_{\underline{\chi}_i})^{n_{\underline{\chi}_i}} \ket{\Omega_F} \;.
\end{equation}
The number operator for a single mode, $N_{\underline{\chi}} = \varphi^*_{\underline{\chi}} \varphi_{\underline{\chi}}$, counts its occupation number, $N_{\underline{\chi}_j}\ket{\{n_{\underline{\chi}_i}\}} = n_{\underline{\chi}_j} \ket{\{n_{\underline{\chi}_i}\}}$. The total number operator, $N = \sum_{\underline{\chi}} N_{\underline{\chi}}$, naturally counts the total number of particles in a given state, $N\ket{\{n_{\underline{\chi}_i}\}} = \left(\sum_{j} n_{\underline{\chi}_j} \right) \ket{\{n_{\underline{\chi}_i}\}}$. \\


\textbf{Operator algebra.} The 2nd quantised operators (see \cite{oritifock} for more details, especially in the context of LQG) are elements of the unital *-algebra generated by $\{\varphi,\varphi^*,I\}$, where $I$ is the identity operator on $\mathcal{H}_F$. These elements are in general polynomial functions $\mathcal{O}(\varphi,\varphi^*,I)$ of the three. We denote this algebra by $\mathcal{A}_F$ which acts on $\mathcal{H}_F$. \\

Operator norm is not defined on $\mathcal{A}_F$ because of the unboundedness of the bosonic ladder operators. As is standard in algebraic treatments of many-body quantum systems, we instead work with exponentiated versions of these which results in a unital C*-algebra, the Weyl algebra.


\subsection{Weyl formulation of GFT algebra} \label{weylGFT}
 
The following Weyl reformulation of the GFT system is based on the well-known literature on algebraic quantum field theory \cite{robinson, sewellbook} surrounding the Fock representation of many-body, non-relativistic systems, suitably adapted to define a framework for GFT required to do statistical mechanics. This formulation has been defined and developed in more detail in \cite{kegeles2017}.\\

The fields $\varphi$ and $\varphi^*$ are operator-valued distributions. The corresponding operators are defined by smearing them with test (wave) functions like so,
\begin{equation}
\varphi(f) := \int_{G^d \times \mathbb{R}^n} d\underline{g} \, d\underline{\phi} \; \overline{f}(\underline{g},\underline{\phi}) \varphi(\underline{g},\underline{\phi}) \;\;\;\;,\;\;\;\; \varphi^*(f) := \int_{G^d \times \mathbb{R}^n} d\underline{g} \, d\underline{\phi} \; f(\underline{g},\underline{\phi}) \varphi^*(\underline{g},\underline{\phi}) 
\end{equation}
where $f \in C^\infty(\mathcal{D}) \overset{\text{dense}}{\subset} L^2(\mathcal{D})$, and $\mathcal{D} \subset G^d\times \mathbb{R}^n$ is a compact region within the full base space that includes the identities. $C^\infty(\mathcal{D})$ is the dense subspace of smooth $L^2$ functions defined on $\mathcal{D}$. Then the commutation relations \eqref{ladderCR} take the following form,
\begin{equation} \label{weylCR}
[\varphi(f_1),\varphi^*(f_2)] = (f_1,f_2) \;\;,\;\; [\varphi(f_1),\varphi(f_2)] = [\varphi^*(f_1),\varphi^*(f_2)] = 0 
\end{equation}
where $(f_1,f_2) = \int_{\mathcal{D}} d\underline{g}\, d\underline{\phi} \;\; \overline{f_1}(\underline{g},\underline{\phi}) f_2(\underline{g},\underline{\phi})$ is the $L^2$-inner product.\\

The choice of the space of test functions usually depends on the situation at hand. We choose it to be the space of those functions in the single-particle $L^2$ space which are smooth and have compact support. In fact, we restrict our single-particle Hilbert space to only smooth $L^2$ functions with compact support, which are thus well-defined in this simple sense (see \cite{kegeles2017} for more details). So from here on, we take as the single-particle Hilbert space $\mathcal{H} = C^\infty(\mathcal{D})$, which is also the space of test functions. An alternative choice would be the widely-used space of Schwartz functions. Another standard choice is to use the space of solutions of a specific model as the space of test functions. \\

Bosonic ladder operators are unbounded in the operator norm on $\mathcal{H}_F$ (and thus are only densely defined). Therefore let's define hermitian operators (in the common dense domain of $\varphi$ and $\varphi^*$), $\Phi(f) := \frac{1}{\sqrt{2}}(\varphi(f) + \varphi^*(f))$, and $\Pi(f) := \frac{1}{i\sqrt{2}}(\varphi(f)-\varphi^*(f))$\footnote{Since $\Pi(f) = \Phi(if)$, both $\varphi$ and $\varphi^*$ can be recovered from $\Phi$ alone, which are the generators of this representation.}$^,$\footnote{In terms of operators $\Phi$ and $\Pi$, the commutation relations take the form, $[\Phi(f_1),\Pi(f_2)] = i\,\text{Re}(f_1,f_2) \;,\; [\Phi(f_1),\Phi(f_2)] = [\Pi(f_1),\Pi(f_2)] = i\,\text{Im}(f_1,f_2)$.}, and consider their exponentiations, $W_F(f) := e^{i\Phi(f)}$ (and $\widetilde{W}_F(f) = e^{i\Pi(f)}$), which:

\begin{itemize}
\item are unitary, $W_F(f)^* = W_F(f)^{-1} = W_F(-f)$ , and

\item satisfy Weyl relations, $W_F(f_1)W_F(f_2) = e^{-\frac{i}{2}\text{Im}(f_1,f_2)}W_F(f_1 + f_2)$.
\end{itemize}

That is, $\{W_F(f)\}_{f \in \mathcal{H}}$ defines a Weyl system \cite{chaiken, robinson} over $\mathcal{H}$ in $\mathcal{H}_F$. It defines a unitary representation of the GFT CR algebra in the Fock space with generators $\Phi$. \\

Retaining this algebraic structure and forgetting (for now) the generators $\Phi$ which lend the concrete representation, an \emph{abstract} bosonic GFT system can be defined by the pair $(\mathcal{A},\mathcal{S})$, where $\mathcal{A}$ is the Weyl algebra\footnote{Fermionic statistics would correspond to a Clifford algebra.} generated by Weyl unitaries $\{W(f) \;|\; f \in \mathcal{H}\}$ and $\mathcal{S}$ is the space of algebraic states (complex-valued, linear, normalised, positive functionals on the algebra) over it. The defining relations of this algebra are, 
\begin{equation} \label{weylCR}
W(f_1)W(f_2) = W(f_2)W(f_1) \, e^{-i \, \text{Im}(f_1,f_2)} = W(f_1+f_2) \, e^{-\frac{i}{2} \text{Im}(f_1,f_2)}\end{equation}

where $f_1,f_2 \in \mathcal{H}$, identity is $\mathcal{I}=W(0)$, and unitarity is $W(f)^{-1} = W(f)^* = W(-f)$. This is a unital C*-algebra, equipped with the C*-norm. The benefits of defining a quantum GFT system with an abstract Weyl algebra stem from the fact that some general results can be deduced, which are representation-independent (and would apply, for example, also to condensate representations, should these turn out to exist, and be unitary inequivalent to the Fock one \cite{kegeles2017}). This allows in particular for exploring structural symmetries at the level of the algebra formulated in terms of automorphisms. In this paper, we will consider examples of automorphisms of $\mathcal{A}$ corresponding to structural symmetries (translations) of the underlying theory. The KMS condition with respect to these automorphisms then leads to the definition of structural equilibrium states which encode stability with respect to the corresponding internal flows of the transformation under consideration.  \\

The Fock system is now generated as the Gelfand-Naimark-Segal (GNS) representation $(\pi_F,\mathcal{H}_F,\Omega_F)$ of the regular Gaussian algebraic state given by $\omega_F[W(f)]:= e^{-\frac{||f||^2}{4}}$. Here $\mathcal{H}_F$ is the GNS representation space which is identical to the one that we constructed in the previous section directly using the ladder operators, via the following identities $\pi_F(W(f)) = W_F(f) = e^{i\Phi(f)}$, for all $W(f) \in \mathcal{A}$. The vector state $\Omega_F$ is the cyclic GNS vacuum generating the representation space, $\pi_F(\mathcal{A})\ket{\Omega_F} \overset{\text{dense}}{\subset} \mathcal{H}_F$. It is the same Fock vacuum that was introduced earlier, the no-space state. Now, for an irreducible representation, which $\pi_F$ is, the bicommutant $\pi_F(\mathcal{A})''$ is the full $\mathcal{B}(\mathcal{H}_F)$, the set of bounded linear operators on $\mathcal{H}_F$. This is simply because the commutant is $\mathbb{C}I$ by irreducibility. By the bicommutant theorem, $\pi_F(\mathcal{A})''$ is also the weak and strong closure of $\pi_F(\mathcal{A})$ in the respective Hilbert space topologies. Instead of working directly with the C*-algebra $\pi_F(\mathcal{A})$, we choose to work with its closure $\mathcal{B}(\mathcal{H}_F)$\footnote{Those elements of $\mathcal{A}_F$ which are (made) bounded in the operator norm are then automatically elements of $\mathcal{B}(\mathcal{H}_F)$.}. Thus our kinematic system is defined by a pair consisting of a von Neumann algebra and the relevant space of states over it, $(\mathcal{B}(\mathcal{H}_F), \mathcal{S}_n)$. Here $\mathcal{S}_n$ is the set of algebraic states $\omega_\rho$ induced by density operators on $\mathcal{H}_F$, given by $\omega_\rho[A] = \Tr(\rho \, A)$ for all $A \in \mathcal{B}(\mathcal{H}_F)$. Since $\pi_F$ is faithful, its normal folium $\mathcal{S}_n$ is dense in the space $\mathcal{S}$ of all states, and therefore is the relevant state space to consider.


\subsubsection{Translation automorphisms}

In section \ref{kmsgibbs} we shall construct Gibbs states which are at equilibrium with respect to translations of the system along the directions of the base manifold. The definitions and constructions relevant for this are presented below. \\

\textbf{$\mathbb{R}^n$-translations.} The natural translation map on $n$ copies of the real line,
\begin{align*}
T_{\underline{\phi}} &: G^d \times \mathbb{R}^n \rightarrow G^d \times \mathbb{R}^n \\
&:(\underline{g},\underline{\phi}') \mapsto (\underline{g},\underline{\phi}'+\underline{\phi})
\end{align*}
induces a complex linear map on $\mathcal{H}$ as a shift of functions to the right, 
\begin{align*}
T^*_{\underline{\phi}} &: \mathcal{H} \rightarrow \mathcal{H} \\ 
&: f(\underline{g},\underline{\phi}') \mapsto (T^*_{\underline{\phi}} f)(\underline{g},\underline{\phi}') := (f\circ T_{-\underline{\phi}})(\underline{g},\underline{\phi}') \,.
\end{align*} 

Notice that $T^*_{\underline{\phi}}$ preserves the $L^2$-space inner product due to translation invariance of the Lebesgue measure, $(T^*_{\underline{\phi}} f_1, T^*_{\underline{\phi}} f_2)_\mathcal{H} = (f_1,f_2)_\mathcal{H} \;$. Let us define a linear map on the Weyl algebra via the action on its set of generators, 
\begin{equation}
\alpha_{\underline{\phi}} : \mathcal{A} \rightarrow \mathcal{A} : W(f) \mapsto W(T^*_{\underline{\phi}}f) \,.
\end{equation}
It is easy to show that for each $\phi$, the map $\alpha_\phi$ defines a *-automorphism of $\mathcal{A}$. The set of maps $\{\alpha_\phi\}_{\phi \in \mathbb{R}}$ forms a 1-parameter group. This defines a representation of the group $\mathbb{R}$ in the group of automorphisms of the algebra, Aut$(\mathcal{A})$. That is, the map $\alpha : \mathbb{R} \rightarrow \text{Aut}(\mathcal{A}) : \phi \mapsto \alpha_\phi$  preserves the algebraic structure of reals, $\alpha_{\phi_1 + \phi_2} = \alpha_{\phi_1}\alpha_{\phi_2}$. Extending this to $\mathbb{R}^n$, the maps $\{\alpha_{\underline{\phi}}\}_{\underline{\phi}\in\mathbb{R}^n}$ now form an $n$-parameter group, and $\alpha$ defines a representation of $\mathbb{R}^n$ in Aut$(\mathcal{A})$. \bigskip


\textbf{$G^d$-left translations.} The natural left translations on a group manifold are diffeomorphisms from $G$ to itself. On $G^d$, it is given by the smooth map,
\begin{equation}
L_{\underline{g}} : (\underline{g}', \underline{\phi}) \mapsto (\underline{g}.\underline{g}', \underline{\phi}) \equiv (g_1g'_1,...,g_dg'_d, \underline{\phi}) \,,
\end{equation}
which induces a map on the space of functions,
\begin{equation}
L^*_{\underline{g}}f(\underline{g}',\underline{\phi}) := (f \circ L_{\underline{g}^{-1}}) (\underline{g}',\underline{\phi}) \,.
\end{equation}
This is the standard left regular, unitary representation of $G$ on $L^2(G)$, extended to the case of $d$ copies of $G$. Thus, $L^*_{\underline{g}}$ also preserves the $L^2$-inner product. With this let us define a linear transformation on the Weyl generators by, 
\begin{equation}
\alpha_{\underline{g}} (W(f)) := W(L^*_{\underline{g}}f)\,.
\end{equation}
Then, map $\alpha_{\underline{g}}$ defines a *-automorphism of $\mathcal{A}$. Also like for $\mathbb{R}^n$-translations, $\alpha: G^d \rightarrow \text{Aut}(\mathcal{A})$ is a representation of $G^d$ in the group of all automorphisms of the algebra as it preserves the algebraic structure, $\alpha_{\underline{g}.\underline{g}'} = \alpha_{\underline{g}}\alpha_{\underline{g}'}\;$. Analogous statements hold for right translations.


\subsubsection{Unitary translations} \label{unis}

Using the following known structural properties of GNS representation spaces \cite{robinson}, the automorphisms defined above can be implemented by unitary transformations in the Fock space as follows. \\

\textbf{$\alpha$-Invariant state.} Let $\omega$ be an $\alpha$-invariant state, i.e. $\omega[\alpha A] = \omega[A]$ for all $A \in \mathcal{A}$, for some $\alpha \in \text{Aut}(\mathcal{A})$. Then, $\alpha$ is implemented by unitary operators $U_\omega$ in the GNS representation space $(\pi_\omega, \mathcal{H}_\omega, \Omega_\omega)$, defined by $U_\omega \, \pi_\omega(A) \, U_\omega^* = \pi_\omega(\alpha A)$ with invariance of the GNS vacuum $U_\omega\Omega_\omega = \Omega_\omega$. Similarly, for the general case when $\omega$ is invariant under a group of automorphisms, then $\alpha(G)$ is implemented by a unitary representation $U_\omega(G)$ of $G$ in $\mathcal{H}_\omega$, such that 
\begin{equation}\label{uni}
U_\omega(g) \, \pi_\omega(A) \, U_\omega^*(g) = \pi_\omega(\alpha_g A) \hspace{0,5cm}\text{with}\hspace{0,5cm} U_\omega(g)\Omega_\omega = \Omega_\omega \,.
\end{equation}

\textbf{Fock state.} We recall that the algebraic Fock state over $\mathcal{A}$ is $ \; \omega_F[W(f)] = e^{-||f||^2/4}, \,$ with the associated GNS representation $(\pi_F, \mathcal{H}_F, \Omega_F)$. Then, any automorphism on $\mathcal{A}$ that is defined via a norm-preserving transformation on $\mathcal{H}$ will leave $\omega_F$ invariant. Thus $\omega_F$ is invariant under the class of norm-preserving transformations of the single-particle $L^2$-space $\mathcal{H}$, including the translation automorphisms of the base manifold as defined above. Therefore, automorphisms $\alpha_{\underline{\phi}}$ and $\alpha_{\underline{g}}$ are implemented by groups of unitary operators in $\mathcal{H}_F$. \\

From the unitary transformations \eqref{uni} as applied to Weyl generators in Fock representation, it is straightforward to see that the GFT ladder operators transform in the familiar way, 
\begin{align*}
U_F(\underline{\phi}') \varphi^{\#}(\underline{g},\underline{\phi}) U_F(\underline{\phi}')^{-1} &= \varphi^{\#}(\underline{g},\underline{\phi}+\underline{\phi}') \\
U_F(\underline{g}') \varphi^{\#}(\underline{g},\underline{\phi}) U_F(\underline{g}')^{-1} &= \varphi^{\#}(\underline{g}'\underline{g},\underline{\phi}) 
\end{align*}
where, $\varphi^\#$ denotes both $\varphi, \varphi^*$. From here on the subscript $F$ on the unitary implementations of the automorphisms will be dropped with the understanding that in this paper $U$ refers only to the unitary representation of some group $G$ in target space $\mathcal{U}(\mathcal{H}_F)$, the group of unitary operators on Fock space. It is important to note that the corresponding group homomorphism $U : G \to \mathcal{U}(\mathcal{H}_F)$, for a group $G$, including several copies of it, is strongly continuous in the Fock space. See appendix \ref{continuityproof} for details.  \\

These transformations defined for $\pi_F(\mathcal{A})$ being bounded can be extended to the von Neumann system $\mathcal{B}(\mathcal{H}_F)$. Therefore, our system $(\mathcal{B}(\mathcal{H}_F), \mathcal{S}_n)$ is now equipped with strongly continuous groups of unitary operators that implement, in the Fock system, internal shifts of the underlying manifold $G^d \times \mathbb{R}^n$.


\section{Structural statistical equilibrium} \label{gfteqm}

Equipped with $(\mathcal{B}(\mathcal{H}_F), \mathcal{S}_n)$ as the kinematic description of a system of an arbitrarily large (but finite) number of GFT quanta, it is straightforward to see that it defines a quantum statistical mechanics for GFT. Let $\omega_\rho \in \mathcal{S}_n$, then the quantities $\omega_\rho[A] = \Tr(\rho A)$, where $A \in \mathcal{B}(\mathcal{H}_F)$ and $\rho$ is a density (trace-class and positive) operator, are the quantum statistical averages of $A$. If additionally, $A$ is self-adjoint and has a structure that admits an appropriate (geometric) interpretation, then $\omega_\rho[A]$ are ensemble averages of observables $A$. $\omega_\rho$ is a statistical mixture of quantum states of particles encoding gravitational and scalar matter degrees of freedom associated with the GFT field. Thus, rewriting of the spin network degrees of freedom within a GFT Fock space allows us to define a statistical mechanics for them.\footnote{Like in conventional statistical mechanics of finite systems, this description would not be expected to have enough structure to support different inequivalent phases. In order to access the phase structure of the theory, one needs to invoke techniques from algebraic statistical mechanics, or non-perturbative renormalization, in GFT and study the thermodynamic limit \cite{kegeles2017}.} \\

Given a density operator $\rho$, then one of the most fundamental of thermodynamic potentials, the partition function $Z$, can be defined as being the normalisation of the statistical distribution at hand. From $Z$ can be defined the thermodynamic free energy $F \propto -\ln Z$. Entropy a la von Neumann is $S = -\Tr(\rho \ln \rho)$. These thermodynamic variables are those whose construction does not really rely on the context in which the statistical mechanical framework is formulated. A macrostate of the system is characterised by a set of thermodynamic variables. This includes $Z, S$ and $F$ (and others derived from them, say via Legendre transforms). The remaining \emph{relevant} macrostate variables need to first be identified depending on the specific system at hand. Then, the complete set of such potentials characterises the macrostate of the system $\rho$, whose microstates are naturally the quantum states contributing to the statistical mixture. Having done so at a formal level, of course, the remaining task would be to identify a suitable physical interpretation for them\footnote{For example, if a thermodynamic volume potential can be defined in analogy with usual QFTs, this would refer to the domain manifold of the GFT fields, i.e. the group manifolds, and would not be immediately related to spatial volumes, as deduced for example by the quantum operator we use below, that is motivated by the quantum geometric interpretation of GFT quanta.}. Investigating the thermodynamics of a GFT system is left to future work. \\

The important class of normal states/density operators that are of interest here are the Gibbs states. In the study of bulk properties of a system of many discrete constituents, these states provide the simplest description of the system, that of equilibrium. Generic Gibbs states can be written as $e^{-\sum_l \beta_l \mathcal{O}_l}$, where $\mathcal{O}_l$ are operators that are of interest in the situation at hand whose state averages $\langle \mathcal{O}_l \rangle$ are fixed, and $\beta_l$ are the corresponding intensive parameters that characterise the equilibrium configuration.\\

In this section we study the structural Gibbs states. As discussed in section \ref{backgroundindependence}, in a background independent context, such a state can be derived from maximisation of entropy or via the KMS condition given some flow (or, in cases where the two are equivalent, both). In the following, we present examples of structural thermodynamical and dynamical Gibbs states within the GFT framework.


\subsection{Equilibrium in geometric volume} \label{sectionvolgibbs}

The volume operator plays a crucial role in quantum gravity. In LQG there exist several different proposals (see \cite{bianchi2011, haggardthesis, *smolin95, *ashtekarlewandowski} and references therein), but in each its spectral values are attached to intertwiners associated with the spin network vertices. Since the GFT Fock space is a rewriting of the same degrees of freedom, here a volume eigenvalue is associated to a Fock quantum. As a first investigation to check the usefulness of this statistical mechanical framework for spin network states, we consider a Gibbs state with respect to a volume operator, $V$, defined on $\mathcal{H}_F$. 


\subsubsection{Volume operator} \label{volumeop}

Here, the scalar matter degrees of freedom taking values in $\mathbb{R}^n$ are neglected. The main reason for this choice is that the volume of a quantum of space is a geometric quantity expected to depend primarily on the group representation data $\underline{\chi}$.\footnote{This is not to say that the volume of the corresponding emergent spacetime manifold that is being modelled discretely would not depend on matter, which it of course does according to GR. With this in mind, the choice of independence of volume from matter degrees of freedom should be viewed as a first step that is simple enough to investigate geometric properties of a theory of fundamental discrete constituents of spacetime. This is also the choice that is commonly made in LQG.} From this perspective, given a multi-particle spin network state, the total volume operator should basically count the number of particles in each mode (defined by the geometric data) and multiply this number by the volume eigenvalue associated to a single particle in that mode. It is then evident that it is an extensive, one-body operator. By extensive we mean that it is proportional to the size of the system (the total number of particles, or spin network nodes); and, by one-body we mean that its total action on any multi-particle state is additive with irreducible contributions from individual actions on a single particle. This additivity is motivated from the notion of attaching a quantum of volume to a spin network node. It is thus valid for all quantum states of the GFT Fock space as long as this perspective of attaching a grain of space to a spin network node holds. This is a direct consequence of the Fock space structure of the Hilbert space, and the associated ladder operators. A typical example of an extensive, one-body operator in a standard many-body quantum system is the total kinetic energy. \\

In the occupation number basis (section \ref{fockspace}) the volume operator has the following form,
\begin{equation} \label{vol}
V = \sum_{\underline{\chi}} v_{\underline{\chi}} \, \varphi^*_{\underline{\chi}} \varphi_{\underline{\chi}} \;\;\;,\;\;\;V\ket{\{n_{\underline{\chi}_i}\}} = \left(\sum_{j} v_{\underline{\chi}_j} n_{\underline{\chi}_j} \right) \ket{\{n_{\underline{\chi}_i}\}} 
\end{equation}

where, mode $\underline{\chi}$ denotes the irreducible representation data associated with an open spin network vertex in the spin basis. Keeping in mind our interpretation of $v_{\underline{\chi}}$ as the volume of a quantum polyhedron with faces coloured by $\underline{\chi}$, the following reasonable assumptions are made. First, the single-mode spectrum is chosen to be real and positive, $v_{\underline{\chi}} \in \mathbb{R}_{> 0}$ (for all $\underline{\chi}$). This implies that the spectrum of the total volume operator $V$ is real and non-negative because it is simply a result of scaling $v_{\underline{\chi}}$ with non-negative natural numbers $n_{\underline{\chi}} \geq 0$. Therefore by construction, $V$ is a positive, self-adjoint element of $\mathcal{B}(\mathcal{H}_F)$. Positivity ensures boundedness and therefore the existence of at least one ground state.  Second, we require uniqueness of the single-particle ground state, i.e. $v_{\underline{\chi}} = v_0 \equiv \text{min}\{v_{\underline{\chi}}\} \Leftrightarrow \underline{\chi} = \underline{\chi}_0$, where $V\ket{\underline{\chi}_0} = v_0 \ket{\underline{\chi}_0}$. Notice that the uniqueness assumption would fail if the degenerate zero eigenvalue for $v_{\underline{\chi}}$ is included in the spectrum, because a zero eigenvalue would correspond to several  different spin configurations.\footnote{We are thankful to Mingyi Zhang for pointing this out.} The exact value of $v_0$ depends on the specifics of the spectrum $v_{\underline{\chi}}$ which in turn depends on the specific quantisation scheme used to define the operator. We stress however that the same uniqueness is assumed only for simplicity, and the following calculations, as well as the definition of the Gibbs state, could be adapted to the situation in which it does not hold. 


\subsubsection{Volume Gibbs state}

Let us consider the following mixed state defined on the GFT Fock space, 
\begin{equation}\label{volumerho}
\rho = \frac{1}{Z} \, e^{-\beta (V-\mu N)} \;\;,\;\; Z = \Tr(e^{-\beta (V - \mu N)})
\end{equation}
with $\beta$ and $\mu$ free real parameters, and $0 < \beta < \infty$. Then, $\rho$ is a well-defined element of $\mathcal{S}_n$ for $\mu \leq v_0$. See appendix \ref{volumechecks} for details on verification of $\rho$ as a genuine density operator on $\mathcal{H}_F$. \\

What does it mean to define such a Gibbs state as generated by the volume operator? Referring back to the discussion in section \ref{backgroundindependence}, specifically to thermodynamical Gibbs states in a background independent setting, a state like \eqref{volumerho} can be best understood as arising from Jaynes' principle of maximisation of entropy, $S = - \langle \ln \rho \rangle$ of the system, under the constraints $\langle I \rangle = 1$, $\langle V \rangle = \bar{V}$ and $\langle N \rangle = \bar{N}$, without any need of a flow. Parameters $\beta$ and $\mu$ enter formally as Lagrange multipliers. From a purely statistical point of view, the corresponding physical picture, intuitively, would be that of a system in contact with a bath, which exchanges quantities corresponding to the operators $V$ and $N$.\footnote{In the case at hand, exchange of particles inevitably leads to exchange of volume (and vice-versa), because the particles themselves carry the quanta of volume. In fact, $\mu$ is just like a constant shift in the volume spectrum, in this example.}$^,$\footnote{For instance, consider a system in a pure (entangled) spin network state with a fixed large number of nodes. Physically, this can be understood as corresponding to a region of space with a fixed total volume. Then, partial tracing over a part of it (the bath) would expectedly result in a (reduced) mixed state for the complementary subsystem. A precise characterisation of the bath and its coupled subsystem, particularly, boundary effects due to spin network edges puncturing the boundary of the subsystem, effects of entanglement across the boundary surface, and the exact conditions for thermalisation of the subsystem to a Gibbs state are outside the scope of the current work.} The macroscopic description of the system is then given by the averages $\bar{V},\bar{N}$ along with the intensive parameters $\beta, \mu$\footnote{Formally $\beta$ and $\mu$ parametrise the class of Gibbs states \eqref{volumerho}. Presently no attempt is made to attach any additional interpretations to them.} which characterise the equilibrium configuration. \\

Given a quantum statistical mechanical framework, defining a state like \eqref{volumerho} is justified from the statistical point of view as stated above. But in the context of quantum gravity, why would such a state be interesting to look at? As hinted at in \cite{montesinosrovelli}, a mixed state generated by geometric operators like volume and area would be expected to describe better the physical state of a region of space rather than an arbitrary pure spin network state, wherein the corresponding macroscopic volume and area of the region are given by statistical averages, $\langle V \rangle_\rho$ and $\langle A \rangle_\rho$, characterising (at least partially) the geometric macrostate. In other studies for example, a similar perspective is held with the aim of defining `geometric' entropies, with respect to area measurements of boundary spin network states in \cite{krasnov96}, volume measurements of bulk spin network states in \cite{astuti2016}, and in several LQG-inspired analyses of quantum black holes microstates \cite{DiazPolo:2011np}. Within the framework described here, such geometrical entropies arise naturally as the von Neumann entropy of a statistical state $\rho$ of geometric observables. \\

In addition to the link that such states could provide to macroscopic geometric observables, we are now going to show that they could be employed to extract an interesting phase purely as a result of the collective behaviour of the constituent fundamental quanta. In the next section we show that when diagonalised in the occupation number basis, the volume Gibbs state as defined in \eqref{volumerho} admits a condensed phase that is populated majorly by GFT quanta in the lowest possible spin configuration $\underline{\chi}_0$. We also comment on a special sub-class of such condensates, the commonly encountered spin-1/2 phase which is a collection of a large number of non-interacting isotropic\footnote{All links incident on an isotropic node are labelled by the same spin.} SU(2) spin network nodes with almost all links labelled by $j=1/2$. 


\subsubsection{Bose-Einstein condensation to low-spin phase}

The occupation number basis of $\mathcal{H}_F$, being the eigenbasis of \eqref{vol}, can be utilised to compute the relevant macrostate variables corresponding to the volume Gibbs state. The partition function is evaluated to be,
\begin{equation} 
Z = \sum_{\{n_{\underline{\chi}_i}\}} \bra{{\{n_{\underline{\chi}_i}\}}} \prod_i e^{-\beta ( v_{\underline{\chi}_i} - \mu ) n_{\underline{\chi}_i} } \ket{{\{n_{\underline{\chi}_i}\}}} = \prod_{\underline{\chi}} \frac{1}{1 - e^{-\beta ( v_{\underline{\chi}} - \mu ) } } \;\;.
\end{equation}

This partition function can be immediately identified as having the same form as that of a gas of free non-relativistic bosons with the Gibbs state defined in terms of the non-interacting Hamiltonian (total kinetic energy) \cite{stringaribook}. In our case however, the simplicity of the state (and consequently of the explicit expressions for the potentials) is not a statement about its underlying dynamics, or the result of some controlled approximation of the same. It is true, however, that in the definition we are presently using, we are neglecting any such dynamical ingredients. For the simple case of the volume operator, unless our geometrical perspective of assigning a quantum of volume to an intertwiner changes, the corresponding operator will always be a one-body extensive operator in the GFT Fock space of spin networks, which, under the general conditions stated above, will always lead to a partition function reminiscent of an ideal Bose gas. It is also true, though, that such a partition function would arise for \emph{any} operator which has the general form \eqref{vol} with a real non-negative spectrum. An interesting example is the kinetic part of a GFT action often used in the literature, with a Laplacian term, $S_K = \int d\underline{g} \; \varphi^*(\underline{g}) (-\sum_{I=1}^d \Delta_{g_I} + M^2) \varphi(\underline{g})$. Since the Wigner modes are eigenstates of the Laplacian, in the spin basis this operator takes a form, $S_K = \sum_{\underline{\chi}} (a_{\underline{\chi}} + M^2)\varphi^*_{\underline{\chi}}\varphi_{\underline{\chi}}$, which is analogous to \eqref{vol}. For the simple case of $SU(2)$, $a_{\underline{\chi}} = \sum_{I=1}^d j_I(j_I + 1)$.   \\

The average total number of particles in state \eqref{volumerho} is,
\begin{equation}
\langle N \rangle_\rho = \Tr(\rho N) = \sum_{\underline{\chi}}  \frac{1}{e^{+\beta ( v_{\underline{\chi}} - \mu )}-1 } = - \frac{\partial F}{\partial \mu} \;,
\end{equation}
where, $F$ is the free energy, $F = \langle V - \mu N \rangle_\rho - \beta^{-1}S  = -\frac{1}{\beta}\ln Z $. The ground state term with the smallest eigenvalue $v_0$ contributes the most to $\langle N \rangle_\rho$. As $\mu \rightarrow v_0$, occupation number of the ground state, $N_0 \equiv \langle N_{\underline{\chi}_0} \rangle_{\rho}$ diverges and the system undergoes condensation, resulting in a macroscopic occupation of the single-particle state $\ket{\underline{\chi}_0}$ with volume $v_0$. A low-spin condensate phase thus arises naturally as a quantum statistical mechanical process of a system of fundamental atoms of space. \\

This is just like the standard Bose condensation \cite{stringaribook}. The order parameter can now be directly seen to be the non-zero expectation value of the GFT field operator, i.e. the condensate wavefunction,
\begin{align*}
\langle \varphi(\underline{g}) \rangle_{\rho} &= \langle \, \sum_{\underline{\chi}} \psi_{\underline{\chi}}(\underline{g}) \varphi_{\underline{\chi}} \, \rangle_\rho = \langle \, \psi_{\underline{\chi}_0}(\underline{g})\varphi_{\underline{\chi}_0} + \sum_{\underline{\chi} \neq \underline{\chi}_0} \psi_{\underline{\chi}}(\underline{g}) \varphi_{\underline{\chi}} \,\rangle_\rho \\
&\overset{\mu \rightarrow v_0}{\longrightarrow} \langle \, \psi_{\underline{\chi}_0}(\underline{g}) \sqrt{N_0} \, \rangle_{\Psi_0} = \sqrt{N_0} \,\psi_{\underline{\chi}_0}(\underline{g}) \;.
\end{align*}
 
The single-particle state $\ket{\underline{\chi}_0}$ characterising the condensate corresponds to a set of spin labels encoding ground state data of whichever\footnote{As long as it satisfies the general properties laid out in section \ref{volumeop}.} volume operator is chosen. A special class of such condensates is for the choice of isotropic vertices of $SU(2)$ spin networks and a ground state corresponding to a minimum spin $j_0 = 1/2$. The above then is a mechanism, rooted purely in the quantum statistical mechanics of GFT quanta, for the emergence of a spin-1/2 phase. This bulk configuration has been identified and used repeatedly as the relevant sector in LQG for LQC, and also in GFT condensate cosmology.  \\

Let us digress briefly to place this result in the context of a recent work \cite{gielenemergence}, wherein a similar result is obtained within the framework of GFT cosmology. There are crucial differences in our analysis relative to theirs. (1) The analysis in \cite{gielenemergence} is carried out at the mean field, `hydrodynamic' \cite{oritireview2016} level in terms of the condensate wavefunction of pure coherent states. On the other hand, here the analysis is directly at the level of the microscopic theory of the GFT quanta which then gives rise to a condensate state, which is not \emph{chosen} to be a coherent state but derived to be $\Psi_0$. In fact, the starting point here is a maximally mixed state. (2) The result in \cite{gielenemergence} holds for isotropic vertices in the GFT coherent state (mean field), while here it holds true for generic anisotropic data. (3) And finally, in \cite{gielenemergence} this phase is shown to emerge from the particular wavefunction solution (which is a function of the relational clock $\phi$) in an asymptotic regime of relational evolution, $\phi \rightarrow \pm \infty$. This re-emphasises the facts that their analysis is restricted to: the mean field approximation, where it is reasonable to consider asymptotics in $\phi$, which would be ill-defined in the underlying full quantum theory, if understood as corresponding to a specific GFT state. Furthermore, since their analysis relies crucially on the inclusion of the relational scalar field $\phi$, it would seem that the consequent result is also confined to GFT models coupled to matter. In this sense our result strengthens theirs because here the low-spin phase is shown to emerge already for only gravitational degrees of freedom and as a structural, model-independent feature of a geometric statistical state. In the same respect, our result is actually closer to the one obtained, in absence of scalar field coupling, but still at the mean field level and restricted to isotropic configurations, but including also non-trivial GFT interactions, in \cite{Pithis:2016wzf}. A more direct comparison with \cite{gielenemergence} could be made by generalising the above condensation mechanism to the case of a GFT coupled to a relational clock. This is left to future work. \\

Finally, given the nice properties of operator $V$ which mimic the Hamiltonian of a system of non-interacting bosons in a box \cite{stringaribook}, the result that this system condenses to the single-particle ground state is not surprising. Still, this simple example illustrates the potential of considering collective, statistical features that are inherent in the perspective that spacetime has a microstructure consisting of fundamental, discrete quantum gravity degrees of freedom. It also illustrates the usefulness of the GFT reformulation of spin network degrees of freedom within a Fock space.


\subsection{Equilibrium in internal translations} \label{kmsgibbs}

Now we turn our attention to the Gibbs states which encode equilibrium via the KMS condition under translations of the base manifold $G^d \times \mathbb{R}^n$ as dictated by the unitary groups of operators $U(\underline{\phi})$ and $U(\underline{g})$ on $\mathcal{H}_F$.


\subsubsection{KMS condition and Gibbs states}

In quantum statistical mechanics and thermal field theory alike, the KMS condition has been recognised as characterising equilibrium in terms of stability under a certain automorphism or flow in the algebra. Originally formulated in terms of thermodynamical Green's functions that are characteristic of Gibbs states \cite{kubo, *martinschwinger}, it was later adapted to an algebraic setting wherein its importance for defining equilibrium states under the thermodynamic limit (of taking the system size to infinity whilst keeping the density of gas finite) was made explicit \cite{HHW1967}. \\

There are several equivalent formulations of the  KMS condition \cite{robinson}. The one we use is as follows. Let $F_{AB}(z)$ be a complex function on the complex plane which is analytic in the strip $\{z \in \mathbb{C} \; | \; 0 < \text{Im}z < \beta\}$ and continuous on its boundaries. A state $\omega$ over an algebra $\mathcal{A}$ can be said to define such a function if  $F_{AB}(z) := \omega[A \, \alpha_z(B)]$, where $A,B \in \mathcal{A}$, and $\alpha$ is a 1-parameter group of automorphisms of $\mathcal{A}$, extended here to a complex variable. Then, the same state is said to satisfy the KMS condition if $\omega[A (\alpha_{t+i\beta}B)] = \omega[(\alpha_t B)A]$, i.e. one has periodicity in the boundary values,
\begin{equation*}
F_{AB}(t + i\beta) = \omega[\alpha_t(B) \, A] \,.
\end{equation*}
The state $\omega$ is then called a KMS state. A KMS state automatically satisfies stationarity, $\omega[\alpha A] = \omega[A]$, which captures the simplest notion of equilibrium. As is well-known, this characterisation of equilibrium survives in the limit of infinite system size, whereas the Gibbs description would fail. \\

Before considering specific Gibbs states corresponding to the translation automorphisms of the GFT system, we first show in generality that given an automorphism and its unitary representation in the GFT Hilbert space $\mathcal{H}_F$, the unique KMS state as defined by this automorphism will be a Gibbs state with respect to the generator of the transformation. The proof proceeds in line with the case of quantum statistical mechanics of a finite, non-relativistic system \cite{robinson}, the main conceptual difference being that in the standard case the automorphism of interest is usually the physical time translations, whereas in our case we understand them to be of a more generic nature and certainly not (strictly) related to time translations. \\

Let $\alpha_t$ be a 1-parameter group of automorphisms of the GFT algebra $\mathcal{A}$ that are represented unitarily in $\pi_F$ by the group of operators $U(t)=e^{it\mathcal{G}}$, where $\mathcal{G}$ is a self-adjoint generator. Note that $t \in \mathbb{R}$ is an arbitrary parameter whose interpretation relies on what kind of transformation the automorphism $\alpha_t$ encodes. Consider a normal state in this representation $\omega_\rho[A] := \text{Tr}(\rho \pi_F(A))$ (for all $A \in \mathcal{A}$), satisfying the KMS condition with respect to $\alpha_t$. Then, by definition of the KMS condition, $\omega_\rho[BA] = \omega_\rho[A(\alpha_{i\beta}B)]$ for all $A,B \in \mathcal{A}$, that is,
\begin{align*} 
\text{Tr}(\rho\, \pi_F(B)\pi_F(A)) = \text{Tr}(\rho\, \pi_F(A)\pi_F(\alpha_{i\beta}B)) &= \text{Tr}(\rho\, \pi_F(A)\,e^{-\beta \mathcal{G}}\pi_F(B)e^{\beta \mathcal{G}})    \\
		&= \text{Tr}(e^{-\beta \mathcal{G}}\pi_F(B)e^{\beta \mathcal{G}}\,\rho\, \pi_F(A))
\end{align*}
Notice that the last equality holds true for all $A\in \mathcal{A}$. This implies
$e^{-\beta \mathcal{G}}\pi_F(B)e^{\beta \mathcal{G}}\,\rho = \rho\, \pi_F(B) \;\; \Rightarrow \;\; [e^{\beta \mathcal{G}}\rho \,, \, \pi_F(B)] = 0 \;$ (for all $B \in \mathcal{A}$) $\;\; \Rightarrow \;\; e^{\beta \mathcal{G}}\rho \in \pi_F(\mathcal{A})'$. Now, $\pi_F$ is irreducible, meaning $\pi_F(\mathcal{A})' = \mathbb{C}I$. Therefore, $e^{\beta \mathcal{G}}\rho \propto \mathbb{I} \; \Rightarrow \; \rho \propto e^{-\beta \mathcal{G}}. $ Thus the unique normal KMS state in $(\pi_F,\mathcal{H}_F,\Omega_F)$ is a Gibbs state. $\hfill \square$  \\


Now let us turn to the von Neumann algebra $\pi_F(\mathcal{A})'' = \mathcal{B}(\mathcal{H}_F)$. Like above, given a KMS state $\omega_\rho$ with respect to an automorphism group $\alpha_t$ of $\mathcal{A}$, and the corresponding unitaries $U_t$ in $(\pi_F,\mathcal{H}_F,\Omega_F)$, then in $(\mathcal{B}(\mathcal{H}_F), \mathcal{S}_n)$, existence of $U_t$ is ensured by the BLT theorem. Then by the KMS condition we have as before,
\begin{equation} \nonumber
\text{Tr}(\rho\, B\,A) =  \text{Tr}(\rho\, A\,e^{-\beta \mathcal{G}}\,B\,e^{\beta \mathcal{G}})   = \text{Tr}(e^{-\beta \mathcal{G}}\,B\,e^{\beta \mathcal{G}}\,\rho\, A)
\end{equation} 
for all $A,B \in \mathcal{B}(\mathcal{H}_F)$. Using the same arguments as above, $e^{\beta \mathcal{G}}\rho \propto \mathbb{I} \; \Rightarrow \; \rho \propto e^{-\beta \mathcal{G}}. $ Thus, given a flow of continuous unitary transformations, the unique KMS state in the system $(\mathcal{B}(\mathcal{H}_F), \mathcal{S}_n)$ is a Gibbs state. $\hfill \square$ 


\subsubsection{Momentum Gibbs states}

Let $\mathfrak{G}$ be a connected Lie group\footnote{Any connected Lie group is path-connected because as a smooth manifold it is locally-path-connected. Thus any two points on $\mathfrak{G}$ can be connected by a continuous curve. The natural curves to consider on any Lie group are the 1-parameter groups generated via the exponential map. Notice also that the groups relevant to us, $SL(2,\mathbb{C}), Spin(4), SU(2)$ and $\mathbb{R}$, are all connected and simply connected, so that their direct product groups are also connected spaces.}, and $g_X(t) = \exp(tX)$, for $t \in \mathbb{R}$, $X \in L(\mathfrak{G})$, be a 1-parameter subgroup in $\mathfrak{G}$ satisfying $g_X(0) = e$ and $\frac{dg_X}{dt}\vert_{t=0} = X$. $L(\mathfrak{G})$ is the corresponding Lie algebra. The generators of generic left translation flows, $g_X(t,g_0)=e^{tX}g_0 = L_{e^{tX}}g_0$, are the right-invariant vector fields, $\mathfrak{X}$. The set of all such vector fields is isomorphic to the Lie algebra by right translations on $\mathfrak{G}$, that is $\{R_{g*}X \;|\; X \in L(\mathfrak{G}), g\in \mathfrak{G}\} = \{\mathfrak{X}(g)\}$.  \\

The map $g_X : \mathbb{R} \to \mathfrak{G} : t \mapsto g_X(t)$ is a continuous group homomorphism, preserving additivity of the reals, $g_X(t_1)g_X(t_2) = g_X(t_1 + t_2)$. Now, let $U: \mathfrak{G} \to \mathcal{U}(\mathfrak{H})$ be a strongly continuous unitary representation of the group $\mathfrak{G}$ in a Hilbert space $\mathfrak{H}$, where $\mathcal{U}(\mathfrak{H})$ is the group of unitary operators on $\mathfrak{H}$. Then, the map $U\circ g_X : t \mapsto U(g_X(t))$ is a strongly continuous 1-parameter group of unitary operators in $\mathcal{U}(\mathfrak{H})$. See appendix \ref{cont} for proof of continuity; whereas the group property is straightforward to see by noticing that $U(g_X(t_1)) U(g_X(t_2)) = U(g_X(t_1)g_X(t_2)) = U(g_X(t_1+t_2))$. In terms of $U_X := U \circ g_X$, it takes the expected form,
$U_X(t_1)U_X(t_2) = U_X(t_1 + t_2) \,$.
Applying Stone's theorem to this strongly continuous group of unitary operators leads to the existence of a self-adjoint (not necessarily bounded) generator $\mathcal{G}_X$ defined on $\mathfrak{H}$, such that
\begin{equation}
U_X(t) = e^{-i\mathcal{G}_X t} \;.
\end{equation}
In terms of the anti-hermitian representation $U_*$ of the Lie algebra (which is the differential map induced by the unitary group representation $U$), the $t$-flow in $\mathfrak{H}$ is implemented by
\begin{equation}
U_X(t) = U(\exp(tX)) = \exp(t\,U_*(X))\,.
\end{equation}
Comparing the previous two equations, we arrive at the expression for the self-adjoint generator,
\begin{equation} \label{gen}
\mathcal{G}_X = iU_*(X)\,.
\end{equation}
$\mathcal{G}_X$ implements infinitesimal translations on quantum states in $\mathfrak{H}$, along the direction of the integral flow of $\mathfrak{X}$, and is thus understood as a momentum operator. \\

Let a density operator $\rho$ on $\mathfrak{H}$ satisfy the KMS condition with respect to translations $U_X(t)$. Then as laid out in the previous section, the state must be of the following Gibbs form,
\begin{equation}
\rho_{X} = \frac{1}{Z} e^{-\beta \mathcal{G}_X} \;\;,\;\; Z = \Tr_{\mathfrak{h}}(e^{-\beta \mathcal{G}_X}) \label{GibbsGFT}
\end{equation}
where $\beta$ is the periodicity in the flow parameter $t$. Naturally, in a given situation, the form of $\mathcal{G}_X$ must be (made to be) such that $Z$ is well-defined. This class of Gibbs states is labelled by both the periodicity $\beta$ and  the generating vector $X$. Therefore, this notion of equilibrium has an intrinsic dependence on the curve used to define it.  \\

The constructions up until now will hold independently of whether $\mathfrak{G}$ is abelian or not. The detailed Lie algebra structure determines whether the system retains its equilibrium properties on the entire $\mathfrak{G}$. In other words, it determines whether the system is stable under arbitrary translation perturbations. The state $\rho_X$, as defined by the curve $g_X(t)$, remains invariant under translations to anywhere on $\mathfrak{G}$ if and only if $\mathfrak{G}$ is abelian. Otherwise, the system is at equilibrium \emph{only} along the curve which defines it. To see this, let us perturb a system at identity $e$ in state $\rho_X$, so that it leaves its defining trajectory $g_X(t)$, and reaches another point $h \in \mathfrak{G}$ which is not on $g_X(t)$, i.e. $h \notin \{g_X(t)\;|\;t \in \mathbb{R}\}$. Since $\mathfrak{G}$ is connected, any element of it can in general be written as a product of exponentials. That is, $h = \exp{Y_1}...\exp{Y_\kappa}$ for some $\kappa$, and $Y_1,..,Y_\kappa \in L(\mathfrak{G})$. Left translation by $h$ is implemented in the Hilbert space by the unitary representation,
\begin{equation*}
U(h) = U(\exp{Y_1})...U(\exp{Y_\kappa}) = \exp(U_*(Y_1))...\exp(U_*(Y_\kappa))\;,
\end{equation*}
which acts on the density operator by,
\begin{equation*}
U(h)^{-1} \rho_X U(h) = e^{-U_*(Y_\kappa)}...e^{-U_*(Y_1)} \; e^{- i \beta U_*(X)} \; e^{U_*(Y_1)}...e^{U_*(Y_\kappa)} \;.
\end{equation*}
For non-abelian $\mathfrak{G}$, clearly $[X,Y] \neq 0$ for arbitrary $X,Y \in L(\mathfrak{G})$. Thus in this case, $U(h)^{-1} \rho_X U(h) \neq \rho_X$. On the contrary, for abelian $\mathfrak{G}$, the Lie bracket is zero and equality will hold for arbitrary $h$. So overall, the notion of equilibrium with respect to shifts on $\mathfrak{G}$ is curve-wise, or direction-wise (``direction" being defined on each point of the manifold by the vector field $\mathfrak{X}$). For non-abelian $\mathfrak{G}$, one can only define an equilibrium state \emph{along} a particular direction\footnote{This is the case, for example, for the Unruh effect treated via the Bisognano-Wichmann construction \cite{bisognanowichmann}. In that case, the symmetry group $\mathfrak{G}$ is the Lorentz group, acting on the base space which is a Rindler wedge of the Minkowski spacetime, and the KMS state of the accelerated observer depends on the specific trajectory generated by a 1-parameter flow of boosts, in say $x^1$ direction, taking the form $\Lambda_{k_1}(t) = e^{t a k_1}$, where $k_1$ is the boost generator, and acceleration $a$ parametrises the strength of this boost.} 
. \\

For the GFT Fock system, the Hilbert space $\mathfrak{H}$ is the Fock space $\mathcal{H}_F$, Lie group $\mathfrak{G}$ is the symmetry group under consideration, which is $G^d$ for group translations or $\mathbb{R}^n$ for internal scalar field translations, acting on the base space $G^d \times \mathbb{R}^n$. The strongly continuous unitary groups $U(g)$ are those constructed in section \ref{unis} which implement the translation automorphisms of the GFT Weyl algebra in $\mathcal{H}_F$. The form of the generators in Fock space is,
\begin{equation} \label{fockgen}
U_*(X) := \int_{G^d \times \mathbb{R}^n} d\underline{g}\,d\underline{\phi} \; \varphi^*(\underline{g},\underline{\phi}) \mathcal{L}_{\mathfrak{X}}\varphi (\underline{g},\underline{\phi}) = -\sum_{\mathfrak{f}} \varphi^*(\mathfrak{f}) \varphi(\mathcal{L}_{\mathfrak{X}}\mathfrak{f}) \,.
\end{equation}
Here $\mathfrak{X}$ is the right-invariant vector field on $\mathfrak{G}$ corresponding to the vector $X \in L(\mathfrak{G})$, and related to it by right translations as, $\mathfrak{X}(g) = R_{g*}X$ (for $g \in \mathfrak{G}$). $\mathcal{L}_{\mathfrak{X}}$ denotes the Lie derivative with respect to the vector field $\mathfrak{X}$. $\{\mathfrak{f}\}$ is an orthonormal basis in the space of test functions. The second equality uses: $(\mathcal{L}_{\mathfrak{X}}\varphi^\#)(f) = -\varphi^\#(\mathcal{L}_{\mathfrak{X}}f)$ for compactly supported test functions $f$, and completeness of the basis $\sum_{\mathfrak{f}}\bar{\mathfrak{f}}(\underline{g}',\underline{\phi}')\mathfrak{f}(\underline{g},\underline{\phi}) = \mathbb{I}(\underline{g},\underline{g}')\delta(\underline{\phi}-\underline{\phi}')$. The corresponding KMS equilibrium states for the GFT system, with respect to the given automorphisms, will then take the form \eqref{GibbsGFT}, with ($i$ times) \eqref{fockgen} in the exponent.


\subsubsection{Equilibrium in $\phi$-translations} \label{KMS-phi}

The above construction of KMS states for GFT systems, in terms of translation automorphisms of the GFT algebra, was rather general. As a specific example of the momentum Gibbs states defined above, this section presents those states that are in equilibrium with respect to flows on $\mathbb{R}^n$ part of the GFT configuration space. These are particularly interesting from a physical perspective. First, we anticipate that in light of the interpretation of $\underline{\phi} \equiv (\phi_1,...,\phi_a,...,\phi_n) \in \mathbb{R}^n$ as $n$ number of minimally coupled scalar fields, the corresponding momenta that generate these internal translations are the scalar field momenta, so there is an immediate meaning to the variables. Second, and more important, the scalar values $\phi$ can be used, as in GFT cosmology, as relational clocks, thus their translations can be related rather directly to physical evolution. This is also the reason why we call these $\underline{\phi}$-translations as {\it internal} (beside the fact that they are technically internal automorphism of the GFT algebra), as  will become apparent in the upcoming section \ref{physeqm} where the full GFT system will be deparametrized with respect to one of these scalar fields so that the resulting translations along this field become external to the system, thus defining relational evolution. The momentum of the clock scalar field defined within the reduced system will then also be the clock Hamiltonian (up to a negative sign). \\

The basis of invariant vector fields on $\mathfrak{G} = \mathbb{R}^n$ is $\{\frac{\partial \,\;\;}{\partial \phi^a}\}$ in cartesian coordinates $(\phi^a)$. These are generated by the set of basis vectors of the Lie algebra $\{E_a\}$. The full set of invariant vector fields is then generated by linearity. For a generic tangent vector, $X = \lambda^a E_a$ (sum over repeated index), the corresponding invariant vector field is $\mathfrak{X} = \lambda^a \partial_a$. Then directly for the basis elements, generators \eqref{fockgen} take the simple, familiar form,
\begin{equation}
U_*(E_a) = \int_{G^d \times \mathbb{R}^n} d\underline{g}\,d\underline{\phi} \; \varphi^*(\underline{g},\underline{\phi}) \frac{\partial}{\partial \phi^a} \varphi (\underline{g},\underline{\phi}) = \sum_{\mathfrak{f}} \varphi^*(\mathfrak{f})\partial_a\varphi(\mathfrak{f}) \,.
\end{equation}
It is straightforward to check that, as is required, $U_*(E_a)$ are anti-hermitian (taking boundary terms to vanish, which is compatible with the ladder operators being defined for test functions with compact support). The corresponding basis of momenta $\mathcal{G}_{E_a}$, denoted here by $P_a$, is
\begin{align}
 P_a \equiv \mathcal{G}_{E_a} &=  i \,U_*(E_a) \label{phimom}  \; = \int_{\mathbb{R}^n} \frac{d\underline{p}}{(2\pi)^n} \sum_{\underline{\chi}} \; p_a \, \varphi^*(\underline{\chi},\underline{p}) \, \varphi(\underline{\chi},\underline{p}) \,.
\end{align}
Evidently, $P_a$ are hermitian, satisfying $(\psi_1,P_a \psi_2) = (P_a\psi_1,\psi_2)$, for all $\psi_1,\psi_2 \in D(P_a)$ in its dense domain. More crucially, from its spectral decomposition in the spin-momentum basis it is clear that $P_a$ are self-adjoint. In addition to needing self-adjointness for $P_a$ to be generators of unitary transformation groups, it is also required in order to interpret $P_a$ as the observable momenta of the respective scalar fields $\phi_a$. We remark that operators $P_a$ as constructed here are the same as those used in the GFT cosmology framework, introduced first in \cite{ewing2016} (for the case $a=1$). Infinitesimal linear transformations of the ladder operators are generated in the expected way, $ \partial_{\phi^a} \varphi^\# (\underline{g},\underline{\phi}) = i[P_a, \varphi^\#(\underline{g},\underline{\phi})]$. \\

From their expressions in the spin-momentum basis, it is clear that $P_a$ are extensive and therefore diagonalise in the occupation number basis, 
\begin{equation*}
P_a\ket{\{n_{\underline{\chi}_i,\,\underline{p}_i}\}} = \left( \sum_j p_{a,j} \,n_{\underline{\chi}_j,\,\underline{p}_j} \right) \ket{\{n_{\underline{\chi}_i,\,\underline{p}_i}\}}
\end{equation*}
where $p_{a,j}$ is the $a^{\text{th}}$ component of $\underline{p}_j \equiv (p_1,...,p_a,...,p_n)_j$. Since $p_a \in \mathbb{R}$, the spectrum $ \sum_j p_{a,j} \,n_{\underline{\chi}_j,\,\underline{p}_j}$ includes arbitrary negative eigenvalues even though $n_{\underline{\chi}_i,\,\underline{p}_i} \geq 0$ for all $i$. Thus $P_a$ in general are not positive operators, and moreover, are unbounded from above and below. Then in order to define Gibbs density operators using $P_a$, as we suggest below, some extra conditions will need to be imposed. \\

The grand-canonical Gibbs states\footnote{The canonical Gibbs states are constructed analogously.} that encode equilibrium with respect to internal $\phi$-translations separately along the $n$ cartesian directions are
\begin{equation}
\rho_a = \frac{1}{Z} e^{-\beta (P_a - \mu N)}
\end{equation}
where $\beta$ is the periodicity in $\phi_a$. $\rho_a$ as defined here is positive and trace-class as long as $\beta (P_a - \mu N)$ is a positive operator in $\mathcal{H}_F$, that is, $\beta \sum_i (p_{a,i} - \mu) \,n_{\underline{\chi}_i,\,\underline{p}_i} \geq 0$ in all basis states $\ket{\{n_{\underline{\chi}_i,\,\underline{p}_i}\}}$ (the extra conditions mentioned above amount to ensuring that this property is satisfied, and lead to two cases, one in which each of the factor is positive and the other in which they are each negative). The proof proceeds in analogy with that of the volume Gibbs state detailed in appendix \ref{volumechecks} and is thus not detailed here.

\

The above states provide then an explicit realization of KMS states in a fundamental quantum gravity system. They also amount to a concrete realization of the thermal time hypothesis \cite{rovelli93, connesrovelli}, since they may be understood as defining implicitly a notion of time, via their corresponding automorphism. Of course, much remains to be done to elucidate and analyse in detail their physical meaning and potential applications.


\section{Physical relational statistical equilibrium} \label{physeqm}

We have recalled in the introduction the fundamental difficulties in defining equilibrium in generally covariant systems, due to the absence of preferred time variables. A general strategy to solve those issues, in the description of the dynamics of such systems is to use matter degrees of freedom as relational clocks, under suitable approximations, and recast the general covariant dynamics in terms of a physical Hamiltonian associated with them. We now consider the same general strategy as a way to solve our (related) issue of defining statistical equilibrium states in full quantum gravity. That is, consider the construction of states which are at equilibrium with respect to relational clocks. It turns out, as it could be expected, that the resulting states are closely related to the structural ones defined above in terms of internal translations and KMS conditions.\\

The Gibbs states defined in the previous section that are generated by the scalar field momenta \eqref{phimom} encode equilibrium with respect to \emph{internal} $\phi_a$-translations. These flows are structural, devoid of any physical model-dependent information, in particular a specific choice of dynamics, and so are the resultant equilibrium states. In this section we define those states which are at equilibrium with respect to $\phi$-translations generated by a (model-dependent) clock Hamiltonian encoding relational dynamics, wherein the scalar field takes on the role of an \emph{external} clock time. The resultant relational system will be `canonical' in clock time which now foliates the original system. Lie brackets \eqref{ladderCR} will be replaced with the corresponding equal-clock-time commutation relations, analogous to the equal-time CRs in a non-covariant system. Our interest in such a setup is natural because GFTs lack a preferred choice of an evolution parameter. The kinematical base space $G^d \times \mathbb{R}^n$ has been constructed in a way so as to facilitate a relational description of the system by coupling $n$ scalar fields. However, there are $n$ possible variables to choose from, and none is preferred over the other. Thus by construction GFTs have a multi-fingered relational time structure, in this specific sense.  \\

The way we approach the task of deparametrizing the GFT system, in this work, is the following. We focus first on the classical description of a single GFT quantum, and sketch how deparametrization works at this simple level, assuming that the GFT dynamics amounts to a specific choice of dynamical constraint here. Then, we consider the extension of the same deparametrization procedure for a system of many such GFT \lq particles\rq, assumed as non-interacting. We then consider the  quantisation of the corresponding deparametrized system of GFT particles, arriving at the corresponding quantum multi-particle system. In the resulting canonical system, we define relational equilibrium Gibbs states. We only sketch the relevant steps of the construction, because a good part of them are straightforward, and because a proper definition and analysis of the corresponding mathematical structures, for GFTs, is beyond the scope of this work. For example, the issue of how to deal with dynamical constraints (classical and quantum) in GFT is in part the subject of \cite{kotechainprep}.


\subsection{Deparametrization in classical GFT}
Before moving on to the quantum picture and placing it within the context of the rest of the paper up until now, we begin the investigation for a classical GFT system within the framework of extended phase space and presymplectic mechanics (\cite{relativisticmech, rovellibook} and related works). The power of presymplectic formulation resides in the fact that it is manifestly covariant and does not require non-relativistic concepts like absolute time to describe dynamics. In fact this was the primary motivation in its development, to be able to describe dynamical systems which are generally covariant, or more generally are constrained systems with a set of gauge symmetries and a vanishing canonical Hamiltonian. In GFTs, we face a similar issue of background independence with the associated absence of a preferred evolution parameter. Therefore, the reason for undertaking classical considerations first is to utilise the existing knowledge already well-positioned to be imported to GFT due to common ingredients required in the description of any classical system: configuration space, phase space and a set of constraints including a dynamical (Hamiltonian) constraint. 


\subsubsection{Single-particle system} \label{one}

The extended classical configuration space for the single-particle sector of a GFT system\footnote{By this we mean a classical point particle living on the GFT base space.} is $\mathcal{C}_{\ex} = G^d \times \mathbb{R}^n \ni (g^I,\phi^a)$. The corresponding phase space is $\Gamma_{\ex} = T^*(\mathcal{C}_{\ex}) \cong G^d \times \mathbb{R}^{n} \times L(G^d)^*  \times \mathbb{R}^n \ni (g^I,\phi^a,x_I,p_{\phi^a})$. States and observables are respectively points and smooth functions on $\Gamma_{\ex}$. Statistical states are smooth positive functions on the phase space, normalised with respect to the Liouville measure. The Poisson bracket on the space of observables defines its algebra structure. The symplectic 2-form on $\Gamma_{\ex}$ is
$\omega_{\ex} = \omega_G \,+\, dp_{\phi^a} \wedge d\phi^a $, where $\omega_G$ is the symplectic 2-form on $T^*(G^d)$. Let us assume that the covariant\footnote{By `covariant' we simply mean `not deparametrized', without any relation to diffeomorphisms.} dynamics of this simple system is encoded in a smooth constraint function $C_{\f} : \Gamma_{\text{ex}} \to \mathbb{R}$ (assuming that there are no additional gauge symmetries). This defines a 1-particle GFT system $(\Gamma_{\ex}, \omega_{\ex}, C_{\f})$. \\

What follows is a brief summary of the presymplectic description of this system formulated in direct analogy with standard treatments \cite{relativisticmech, rovellibook}. The vector field $X_{C_{\f}}$ corresponding to the constraint is defined by the equation $\omega_{\ex}(X_{C_{\f}}) = -dC_{\f}$. Constraint surface $\Sigma$ is characterised by $C_{\f} = 0$. The embedding $\imath: \Sigma \to \Gamma_{\ex}$ of the constraint surface in the full phase space induces a presymplectic structure on $\Sigma$ via the pull-back, $\omega_\Sigma = \imath^*\omega_{\ex}$. The null orbits of $\omega_\Sigma$ are the graphs of physical motions encoding unparametrized correlations between the dynamical variables of the theory. These gauge orbits are integral curves of the vector field $X_{C_{\f}}$ satisfying the equations of motion $\omega_\Sigma(X_{C_{\f}}) = 0$. The set of all such orbits is the physical phase space $\Gamma_{\phy}$ that is projected down from $\Sigma$ via a map $\pi: \Sigma \to \Gamma_{\phy}$. $\Gamma_{\phy}$ is equipped with a symplectic 2-form induced from $\Sigma$ by push-forward along the projection map, $\omega_{\phy} = \pi_*\omega_{\Sigma}$. $(\Gamma_{\phy}, \omega_{\phy})$ is the space of solutions of the system and a physical flow means a flow on this space. It is important to notice here that a canonical time or clock structure is still lacking. \\

Deparametrizing this classical system, with respect to, say, the $c^{\text{th}}$ scalar field $\phi^c$, means reducing the full system to one wherein the field $\phi^c$ acts as a good clock. This entails two separate approximations to $C_{\f}$,
\begin{align}
C_{\f} (g^I, \phi^a, x_I, p_{\phi^a}) &\approx p_{\phi^c} + \tilde{C}(g^I, \phi^a, x_I, p_{\phi^\alpha}) \\
&\approx p_{\phi^c} + H(g^I, \phi^\alpha, x_I, p_{\phi^\alpha}) 
\end{align}
where the fixed index $c$ denotes `clock', and index $\alpha \in \{1,2,...,n-1\}$ runs over the remaining scalar field degrees of freedom that are not intended to be used as clocks and remain internal to the system. The first approximation retains terms up to the first order in clock momentum. At this level of approximation the $\tilde{C}$ part is a function of the clock time $\phi^c$. These two features mean that at this level of approximation $\phi^c$ behaves as a clock, but only locally since its momentum is not necessarily conserved in the clock time. Furthermore, by linearising in $p_{\phi^c}$, we have fixed a reference frame defined by the physical matter field $\phi^c$. At the second level, $\tilde{C}$ is approximated by a genuine  Hamiltonian $H$ that is independent of $\phi^c$, so that on-shell we have conservation of the clock momentum $\partial_{\phi^c}p_{\phi^c} = 0$. $H$ generates relational dynamics in $\phi^c$, which now acts as a global clock for this deparametrized system.\footnote{A nice example illustrating these points well is a classical relativistic particle whose covariant dynamics is given by $C_{\f} = p^2 - m^2$. In this case, the complete dynamical, presymplectic description does not require deparametrization, i.e. given the extended phase space along with $C_{\f}$, its constraint surface is well-defined. However, this system is deparametrizable, which means that it is possible to bring $C_{\f}$ to a manifestly non-covariant form without changing the physics (unlike for the general case mentioned above where the two approximations may change the physical content of the system). The full constraint can be rewritten as $C = p_0 + \sqrt{\underline{p}^2 + m^2}$ which now describes the same relativistic particle system but in a fixed Lorentz frame where the configuration variable $x^0$ has been chosen as the the clock variable and the corresponding clock dynamics is in $H = \sqrt{\underline{p}^2 + m^2}$.}  \\

After the above approximations, we have a new system $(\Gamma_{\ex}, \omega_{\ex}, C)$, with 
\begin{equation} \label{deparamC}
C = p_{\phi^c} + H(g^I, \phi^\alpha, x_I, p_{\phi^\alpha})
\end{equation}
deparametrized with respect to one of the extended configuration variables $\phi^c$ which takes on the role of a good global clock. The presymplectic mechanics now takes on a structure mirroring that of a non-relativistic particle in spacetime. The constraint surface defined by the vanishing of the relevant constraint, here $C=0$, now admits the topology of a foliation in clock time, $\Sigma = \mathbb{R} \times \Gamma_{\can} \ni (\phi^c, g^I, \phi^\alpha, x_I, p_{\phi^\alpha})$. This form of $\Sigma$ is a characteristic feature of a system with a clock structure. The reduced, canonical phase space is $\Gamma_{\can} = T^*(\mathcal{C}_{\can}) \ni (g^I, \phi^\alpha, x_I, p_{\phi^\alpha})$ where $\mathcal{C}_{\can} = G^d \times \mathbb{R}^{n-1}$ is the reduced configuration space. The function $H : \Gamma_{\can} \to \mathbb{R}$ is the clock Hamiltonian encoding relational dynamics in $\phi^c$, and one can define the standard symplectic Hamiltonian mechanics with respect to it.


\subsubsection{Multi-particle system} \label{two}
We want now to extend the above deparametrization procedure beyond the one-particle sector of the GFT system. To begin with, let us consider the simplest case of two, non-interacting particles \cite{chircohaggard, chircojosset16}. Let $\Gamma^{(1)} \ni (g^{(1)I},\phi^{(1)a},x^{(1)}_I,p^{(1)}_{\phi^a})$ and $\Gamma^{(2)} \ni (g^{(2)I},\phi^{(2)a},x^{(2)}_I,p^{(2)}_{\phi^a})$ be the extended phase spaces of particles 1 and 2 respectively. Phase space of the composite system is $\Gamma = \Gamma^{(1)} \times \Gamma^{(2)}$ with symplectic 2-form $\omega = \omega^{(1)} + \omega^{(2)}$. Notice that each particle is equipped with $n$ possible clocks. The aim is to select a single global clock for the composite system so as to then be able to define a common equilibrium for the total system.  Let the individual (possibly covariant) dynamics of each particle be given by constraint functions $C^{(1,2)}_{\f} : \Gamma^{(1,2)} \to \mathbb{R}$. Deparametrizing particle 1 with respect to say field $\phi^{(1)c_1}$, and particle 2 with respect to say $\phi^{(2)c_2}$, gives the new constraints for each,
\begin{align}
C^{(1)} &= p^{(1)}_{\phi^{c_1}} + H^{(1)}(g^{(1)I},\phi^{(1)\alpha},x^{(1)}_I,p^{(1)}_{\phi^\alpha})  \approx 0 \nonumber \\
C^{(2)} &= p^{(2)}_{\phi^{c_2}} + H^{(2)}(g^{(2)I},\phi^{(2)\alpha},x^{(2)}_I,p^{(2)}_{\phi^\alpha})  \approx 0 \label{particledep}
\end{align}
where $H^{(1,2)}$ are functions on the individual reduced phase spaces $\Gamma^{(1,2)}_{\can} \ni (g^{(1,2)I},\phi^{(1,2)\alpha},x^{(1,2)}_I,p^{(1,2)}_{\phi^\alpha})$ with symplectic 2-forms $\omega^{(1,2)}_{\can}$. This is a complete theoretical description of the deparametrized 2-particle system. However, it is inconveniently described in terms of two different clocks ascribed to each particle separately. We are seeking a single clock. This system can equivalently be reformulated \cite{chircohaggard} in terms of a single constraint, 
\begin{equation}
C^{(1)} + C^{(2)} \approx 0 
\end{equation}
along with a second-class constraint $C^{(1)}-C^{(2)}$ and the following gauge-fixing condition. Choose $\phi^{(1)c_1} = t$ and $\phi^{(2)c_2} = F(t)$. Then the gauge-fixed 2-form on $\Gamma$ is, 
\begin{align}  
\tilde{\omega}  &=   \omega_{\can} \;+\; dp_t \wedge dt \;, \;\; \text{where} \\
p_t &= p^{(1)}_{\phi^{c_1}} + F'(t)p^{(2)}_{\phi^{c_2}}
\end{align}
is the clock momentum of the single clock $t$ (prime $'$ denoting total derivative with respect to $t$), and $\omega_{\can} = \omega^{(1)}_{\can} + \omega^{(2)}_{\can}$ is the symplectic form on the reduced phase space $\Gamma_{\can} = \Gamma^{(1)}_{\can} \times \Gamma^{(2)}_{\can}$ of the composite system. The first-class constraint can now be rewritten as, $C = p_t + H^{(1)} + F'(t)H^{(2)}$. The Hamiltonian $H^{(1)} + F'(t)H^{(2)}$ is independent of clock $t$ iff $F'(t)=k$, for $k$ an arbitrary non-zero real constant. That is, $t$ is a good clock for the choice of affine gauge $F(t) = kt + \tilde{k}$. This gives,
\begin{equation}
C = p_t + H \,,
\end{equation}
where $H = H^{(1)} + kH^{(2)}$. Now that the 2-particle system has been brought to the form of a standard Hamiltonian system with clock time $t$, the remaining elements for the complete extended symplectic description can be easily identified. The extended configuration space is $\mathcal{C}_{\ex} = \mathbb{R} \times \mathcal{C}_{\can} \ni (t, g^{(1)I}, \phi^{(1)\alpha}, g^{(2)J}, \phi^{(2)\gamma})$. The extended phase space is $\Gamma_{\ex} = T^*(\mathcal{C}_{\ex})$ with $\omega_{\ex} = \tilde{\omega}$. The constraint function $C = p_t + H = 0$ defines the presymplectic surface $\Sigma = \mathbb{R} \times \Gamma_{\can} \ni (t, g^{(1)I},\phi^{(1)\alpha},x^{(1)}_I,p^{(1)}_{\phi^\alpha},g^{(2)J},\phi^{(2)\gamma},x^{(2)}_J,p^{(2)}_{\phi^\gamma})$, with $\omega_\Sigma = \omega_{\can} - dH \wedge dt$. This is a complete description of a non-interacting 2-particle system, with a single relational clock $t$. \\

For an $N$-particle non-interacting system, each with $n$ possible clocks, the extension of the above procedure is direct. Select \emph{any} one $\phi$ variable as a clock for each individual particle (i.e. bring the individual full constraints of each particle to deparametrized forms like in \eqref{particledep}). Then, given one clock per particle, identifying a global clock for all particles means choosing any one at random (call it $t$) and synchronizing the rest with this one via affine functions $F_2(t),...,F_{N}(t)$. This defines a relational system on $\mathcal{C}_{\ex} \ni (t, g^{(1)I}, \phi^{(1)\alpha},...,g^{(N)J}, \phi^{(N)\gamma})$, $\Gamma_{\ex} = T^*(\mathcal{C}_{\ex})$, with constraint function $C = p_t + H$ on $\Gamma_{\ex}$, and Hamiltonian function $H = H^{(1)} + k_2H^{(2)} + ... + k_N H^{(N)}$ on $\Gamma_{\can}$. \\ \\

Before moving on to quantisation, let us pause to make a few important remarks, summarise sections \ref{one} and \ref{two}, and set the notation. The following discussion is meant to clarify: (1) that we are dealing with two different `before' and `after' (deparametrization) systems, one that is covariant and the other that is derived from the first via the deparametrization approximations\footnote{In general the two systems are physically distinct. However, it is possible that deparametrization does not change the physical content of the theory. This corresponds to the case in which the deparametrization steps outlined above do not correspond to approximations of the dynamics, but to exact re-writing or gauge-fixing. Such systems are usually known as deparametrizable. An example is that of a relativistic particle.}; (2) the conceptual and notational differences between these two systems; and (3) that one of these two systems, the deparametrized one, includes within it a canonical system (that is eventually quantised), which is `canonical' with respect to the relational clock that is selected during the process of deparametrization. \\

For the 1-particle system the `before' picture is one wherein the system is fully covariant and the corresponding kinematics consist of the configuration space $\mathcal{C}_{\ex}^{\cov} = G^d \times \mathbb{R}^n \ni (g^I, \phi^a)$ and phase space $\Gamma_{\ex}^{\cov} = T^*(\mathcal{C}_{\ex}^{\cov})$. Covariant dynamics is encoded in a Hamiltonian constraint function $C_{\f}$ on $\Gamma_{\ex}^{\cov}$, which defines a constraint hypersurface in $\Gamma_{\ex}^{\cov}$. The `after' picture defines the second system, which includes the canonical one. The extended configuration space of the deparametrized system is $\mathcal{C}_{\ex}^{\dep} = \mathbb{R} \times (G^d \times \mathbb{R}^{n-1}) \ni (t, g^I, \phi^\alpha)$, where the 1-particle canonical configuration space is $\mathcal{C}_{\can} = G^d \times \mathbb{R}^{n-1}$. Here, we have denoted $\phi^c \equiv t$. Canonical variables are those dynamical variables of the original covariant system $\mathcal{C}_{\ex}^{\cov}$ which are \emph{not} used as clocks. The extended phase space is $\Gamma_{\ex}^{\dep} = T^*(\mathcal{C}_{\ex}^{\dep})$. Deparametrized dynamics is encoded in a constraint function $C$ on $\Gamma_{\ex}^{\dep}$ of the form, $C = p_t + H$, where $H$ is a smooth function on the canonical phase space $\Gamma_{\can} = T^*(\mathcal{C}_{\can})$. It is a genuine Hamiltonian defining dynamical evolution with respect to the relational clock $t$. The constraint surface (satisfying $C=0$) is $\Sigma = \mathbb{R} \times \Gamma_{\can}$, characterised by a foliation consisting of slices $\Gamma_{\can}$ along clock $t$. This form of $\Sigma$ and the existence of the canonical subsystem is a direct consequence of deparametrization. In other words, the canonical subsystem is absent for a generic non-deparametrized, covariant system $(\Gamma_{\ex}^{\cov}, C_{\f})$. As a final remark, note that for the 1-particle system, the covariant and deparametrized kinematic descriptions in the respective configuration spaces are identical. As will be seen below, this does not hold for an $N$-particle system with $N>1$, when seeking a description with a single clock. \\

For the non-interacting $N$-particle system, the `before' system consists of the covariant extended configuration space $\mathcal{C}_{\ex, N}^{\cov} = (G^d \times \mathbb{R}^n)^{\times N} \ni (g^{(1)I}, \phi^{(1)a}, ..., g^{(N)J}, \phi^{(N)b})$ and the associated phase space $\Gamma_{\ex, N}^{\cov} = T^*(\mathcal{C}_{\ex, N}^{\cov}) = (\Gamma_{\ex}^{\cov})^{\times N}$. The covariant dynamics is encoded in a set of Hamiltonian constraints $C_{\f}^{(1)},...,C_{\f}^{(N)}$, each defined on the respective copies of the 1-particle covariant extended phase space $\Gamma_{\ex}^{\cov}$. The `after' system is deparametrized with a single clock $t$. As before, existence of this clock structure means that the extended symplectic description takes on the form of a non-relativistic system. The extended configuration space is $\mathcal{C}_{\ex,N}^{\dep} = \mathbb{R} \times \mathcal{C}_{\can, N}$ where the $N$-particle reduced configuration space is $\mathcal{C}_{\can, N} = (G^d \times \mathbb{R}^{n-1})^{\times N}$, so that $(t, g^{(1)I}, \phi^{(1)\alpha},...,g^{(N)J}, \phi^{(N)\gamma}) \in \mathcal{C}_{\ex,N}^{\dep}$. The extended phase space of the deparametrized system is $\Gamma_{\ex, N}^{\dep} = T^*(\mathcal{C}_{\ex,N}^{\dep})$. The deparametrized dynamics is encoded in a Hamiltonian constraint function, \begin{equation*}
C_N = p_t + H_N
\end{equation*} 
defined on $\Gamma_{\ex,N}^{\dep}$. Constraint surface $\Sigma = \mathbb{R} \times \Gamma_{\can,N}$ is characterised by $C_N=0$. The canonical phase space is $\Gamma_{\can, N} = T^*(\mathcal{C}_{\can, N})$. Relational dynamics is encoded in the clock Hamiltonian defined on $\Gamma_{\can, N}$ given by, 
\begin{equation} \label{hamiltonianN}
H_N = \sum_{i=1}^N k_i H^{(i)} \,,
\end{equation}
for arbitrary real non-zero constants $k_i$ which encode the rates of synchronization between the $N$ different clocks, one per particle. Functions $H^{(i)}$ are single-particle clock Hamiltonians defined on the respective copies of single-particle reduced phase space $\Gamma_{\can}$. We can already anticipate that relational Gibbs states in a multi-particle system with fixed $N$ are those that are stationary with respect to the $t$-flow of a Hamiltonian $H_N$.


\subsection{Quantisation} \label{quantise}

We now move on to the quantisation of the above deparametrized many-body system. Our treatment is again limited to outlining the basic steps, even when they can be made mathematically rigorous, because they are not particularly enlightening in themselves, at least for our present purposes. In the context of deparametrization, since we are primarily interested in scalar degrees of freedom residing in copies of $\mathbb{R}$, we shall be content with omitting rigorous details about the symplectic structure on $T^*(G)$ and its subsequent quantisation to a commutator algebra. We shall also not choose any specific quantisation map and focus only on the general ideas required to eventually define a $\phi$-relational Gibbs operator. More mathematical details can be found in \cite{guedes2013}, including examples of quantisation maps for $T^*(G)$.\\

Quantisation maps the phase space to a Hilbert space, and the classical algebra of observables (including one which encodes the dynamics) as smooth (real) functions on the phase space to (self-adjoint) operators on the Hilbert space with the Poisson bracket on the former being mapped to a commutator bracket in the latter. \\

For the covariant 1-particle system, the phase space $\Gamma_{\ex}^{\cov}=T^*(G^d \times \mathbb{R}^n)$ maps to $\mathcal{H} = L^2(G^d \times \mathbb{R}^n)$, which is the 1-particle Hilbert space that we considered in section \ref{fockspace}. Observables are the algebra of $C^\infty$-functions on $\Gamma_{\ex}^{\cov}$ which map to operators on $\mathcal{H}$, with the Poisson structure on the former being mapped to the Heisenberg algebra on the latter. Specifically, for the matter degrees of freedom, this is $\{\phi^a, p_{\phi^b}\} = \delta_{ab} \rightsquigarrow [\widehat{\phi^a}, \widehat{p_{\phi^b}}] = i\delta_{ab}$ (where the hat denotes some quantisation map). Notice here that all $n$ scalar fields are quantised. This is in contrast with the corresponding case of the canonical system. In this case, the canonical phase space $\Gamma_{\can} = T^*(G^d \times \mathbb{R}^{n-1})$ maps to a canonical Hilbert space $\mathcal{H}_{\can} = L^2(G^d \times \mathbb{R}^{n-1})$, with the algebra again mapping from functions on $\Gamma_{\can}$ to operators on $\mathcal{H}_{\can}$. But now, the brackets defining the algebra structure of the system are reduced by one in number, as a direct consequence of the reduction of the base space by one copy of $\mathbb{R}$ (to which the clock variable belongs). Under quantisation we now have, $\{\phi^\alpha, p_{\phi^\gamma}\} = \delta_{\alpha\gamma} \rightsquigarrow [\widehat{\phi^\alpha}, \widehat{p_{\phi^\gamma}}] = i\delta_{\alpha\gamma}$, where $\alpha, \gamma = 1,...,n-1$. The commutator corresponding initially to the clock $\phi$-variable is now identically zero, that is $[\widehat{\phi^c}, \widehat{p_{\phi^c}}] = 0$, meaning that the corresponding degrees of freedom are treated as entirely classical; moreover, their intrinsic dynamics is neglected. In other words, this quantum canonical system is one in which there exists a classical clock, which was quantum in the original quantum covariant system. Dynamics is defined via a Hamiltonian operator $\hat{H}$ on $\mathcal{H}_{\can}$ giving evolution with respect to the clock. \\

For the covariant $N$-particle system, $\Gamma_{\ex,N}^{\cov} = (\Gamma_{\ex}^{\cov})^{\times N}$ is mapped to $\mathcal{H}_N = \mathcal{H}^{\otimes N}$. Algebra of smooth functions on $\Gamma_{\ex,N}^{\cov}$ is mapped to a *-algebra on $\mathcal{H}_N$, whose quantum matter fields now satisfy $[\widehat{\phi^{(i)a}}, \widehat{p^{(j)}_{\phi^b}}] = i\delta_{ab}\delta_{ij}$, where $i,j=1,2,...,N$ denote the particle label. This is the multi-particle sector as considered in section \ref{fockspace}. Again, we note that none of the $N$ particles have chosen a clock yet, that is all $n \times N$ number of scalar fields are quantum. In the corresponding canonical quantum system $\mathcal{H}_{\can, N}= \mathcal{H}_{\can}^{\otimes N}$ with the algebra of observables on it, the single clock variable $t$ (which is synced with all the separate clocks now carried by each particle) is classical. The Hamiltonian operator defining $t$-evolution, for fixed $N$, is given by the operator $\hat{H}_N = \sum_{i=1}^N k_i \hat{H}^{(i)}$, where $\hat{H}^{(i)}$ are the separate Hamiltonians of each particle scaled by the respective rates at which the different clocks are synchronized. Notice again that we are neglecting interactions between the different particles in $\hat{H}_N$. \\

In the multi-particle case, it is worthwhile to also look at the quantised extended deparametrized system. This consists of $\Gamma_{\ex,N}^{\dep}$ being quantised to $\mathcal{H}_{N}^{\dep} = L^2(\mathbb{R}\times \mathcal{C}_{\can,N})$ and the corresponding Heisenberg algebra has two additional generators (compared to the canonical system) satisfying $[\widehat{t}, \widehat{p_t}] = i$. Such a system is different from both the quantum extended covariant and the quantum canonical. In the former, there is no single clock variable. In the latter, there is one but it is no longer quantum. Quantising the extended deparametrized system is like quantising a non-relativistic particle at the level of its extended phase space, which includes Newtonian time and its conjugate momentum as phase space variables. Quantising Newtonian time to define the corresponding operator comes with its own set of conceptual and technical problems. However our case is fundamentally different because here $t$ is really a function of the physical matter degrees of freedom. Therefore for a multi-particle system, one ends up with three different quantum systems: quantum extended covariant, quantum extended deparametrized and quantum canonical. The last two each have a potential global clock parameter, and going from the former to the latter is the step of making this variable classical and therefore treating it as a perfect, thus idealized, clock. This distinction between quantum extended covariant and extended deparametrized systems is absent in the simple 1-particle system because in this case deparametrization does not require the extra step of syncing the different clocks (as there is only one). It only requires choosing one out of $n$ so that the kinematics of both systems ends up being identical.  \\

We arrive now at the quantum Fock systems built out of the above $N$-particle systems. The covariant Fock system composed of the $N$-particle quantum extended covariant systems as described above is the GFT Fock system as detailed in section \ref{gft} (and used subsequently to construct structural Gibbs states in section \ref{gfteqm}). On the other hand, the canonical Fock system is as follows. The canonical (bosonic) Fock space is 
\begin{equation}
\mathcal{H}_{\can, F} = \bigoplus_{N\geq0} \text{sym}\, \mathcal{H}_{\can}^{\otimes N} \;
\end{equation}
where the 1-particle canonical Hilbert space is $\mathcal{H}_{\can} = L^2(G^d \times \mathbb{R}^{n-1})$. $\mathcal{H}_{\can,F}$ is generated by ladder operators $\hat{\varphi}, \hat{\varphi}^*$ acting on a Fock vacuum, and satisfying the equal- (Fock-) clock-time commutation relations,
\begin{equation} \label{canladderCR}
[\hat{\varphi}(t_F,\underline{g}_1,\vec{\phi}_1),\hat{\varphi}^*(t_F,\underline{g}_2,\vec{\phi}_2)] = \mathbb{I}(\underline{g}_1,\underline{g}_2)\delta(\vec{\phi}_1 - \vec{\phi}_2) \;\;,\;\; [\hat{\varphi}, \hat{\varphi}] = [\hat{\varphi}^*, \hat{\varphi}^*] =0
\end{equation}
where $\mathbb{I}$ and $\delta$ are delta distributions on $G^d$ and $\mathbb{R}^{n-1}$ respectively. Notation $\vec{.}$ has been used to make explicit the difference between variables $\phi$ in canonical and covariant systems. Here $\vec{\phi} \equiv (\phi^1,...,\phi^{n-1})$ denote the canonical variables whereas earlier, $\underline{\phi} \equiv (\phi^1,...,\phi^{n})$ belonged to the covariant system in which all scalar fields were internal variables. $\underline{g}$ continues to denote $(g_1,...,g_d)$. The associated canonical Weyl system is now based on test functions which are defined on the reduced configuration space $\mathcal{C}_{\can} = G^d \times \mathbb{R}^{n-1}$, and analogous constructions to those considered in the beginning of section \ref{weylGFT} follow through. The *-algebra now consists of polynomial functions of the generators $\varphi,\varphi^*$ and $I$ over the reduced base space $G^d \times \mathbb{R}^{n-1}$. For example, the number operator now takes the form, 
\begin{equation*}
\hat{N} = \int_{G^d\times \mathbb{R}^{n-1}} d\underline{g}\,d\vec{\phi} \; \hat{\varphi}^*(\underline{g},\vec{\phi})\hat{\varphi}(\underline{g},\vec{\phi}) \,. \end{equation*}
Note that one can understand these quantities also as observables in the full theory, just computed at given values of the relational clock variable. The heuristic interpretation is valid, but the actual algebraic properties of these observables would be (potentially very) different. \\

Comparing the algebra \eqref{canladderCR} to \eqref{ladderCR}, it is evident that the Fock system of \eqref{canladderCR} describes a canonical setup, but the nature of the time $t_F$ requires clarifications, which we now provide, and more work, which we leave for the future. A canonical Fock system requires a global time variable which is common to all the different multi-particle sectors, that is for a varying $N$. In other words, we are seeking a clock variable, extracted somehow from the original covariant system, which in the reduced canonical system plays a role similar to the time in usual many-body quantum physics. In the case of GFTs this time is a relational variable (or a function of several such variables). To get such a global clock for the canonical Fock system, one needs: 1) a Hamiltonian constraint operator defining some model, since the definition of a relational clock is always model-dependent due to the relational variable itself being one of the dynamical variables of the full system; and 2) the Fock time variable must be accessible from all $N$-particle sectors, i.e. its construction/definition must be compatible with changing the total particle number.  \\

However what we have currently (section \ref{two}) is a procedure of extracting a clock for a given {\it fixed} $N$-particle sector. To see this, consider a simple example. Take a system with two non-interacting particles, each equipped with its own clocks $\phi^{c_1}$ and $\phi^{c_2}$, along with their clock Hamiltonians $H^{(1)}$ and $H^{(2)}$ respectively. Let $t_2$ be the global clock time such that $\phi^{c_1} = F_1(t_2)=k_1t_2 + \tilde{k_1}$ and $\phi^{c_2} = F_2(t_2)=k_2t_2 + \tilde{k_2}$. The $t_2$-clock Hamiltonian is $H_2 = k_1H^{(1)} + k_2H^{(2)}$. Now let's add a third particle to the mix, such that the resultant system remains non-interacting. Then in the new system, the global clock variable $t_3$ will be different from $t_2$, corresponding to a changed relational dynamics given now by $H_3$ which has a non-zero contribution from the dynamics of the third particle. Thus even in the simple case of no interactions, choosing a global time goes hand in hand with choosing a Hamiltonian dynamics restricted to a fixed $N$. Even in an interacting system, with $\hat{H}_N = \hat{H}_{\text{free},N} + \hat{H}_{\text{int},N}$ (where the interaction part spans the configuration space of the different particles simultaneously), there is a preferred clock time $t_N$ corresponding to a given choice of $\hat{H}_N$. Changing the total Hamiltonian to include more particles will also change the corresponding time variable. The case of including interactions in an $N$-particle GFT system will be considered elsewhere. Consequently also the relational Gibbs states constructed in the next section are defined only for a given $N$-particle system, both in classical and quantum cases. \\

Let us make a final remark regarding deparametrization from the perspective of a full quantum covariant theory. The strategy employed here (for a finite dimensional system) is to start from a classical constrained system, deparametrize it to get a classical canonical system with respect to a relational clock, and then quantise the canonical system leaving the clock as classical. A more fundamental, and challenging, construction leading possibly to a more physical sort of (approximate) deparametrization is to begin from the complete, fundamental quantum theory (in our case, the GFT Fock system as detailed in section \ref{gft}) in which all possible relational scalar fields are quantum. Then deparametrizing would mean to identify a relevant regime of the full theory in which one of the coupled scalar fields becomes semi-classical, and {\it only then} apply the deparametrization approximations outlined in the classical case to our full quantum system. For example, such a regime could correspond to restricting the system to a class of semi-classical coherent states with respect to the chosen would-be clock variable. We discuss this line of thought no further and leave it as an interesting future project.


\subsection{Relational equilibrium} \label{releqm}

As we have discussed in the previous sections, in order to deparametrize the system we need to impose the dynamical constraint of the theory. Moreover, deparametrizing, and obtaining a good canonical structure in terms of a relational clock, amounts to the approximation
\begin{equation}
C_{\f} \overset{\text{deparam.}}{\longrightarrow} C = p_t + H_N \;,
\end{equation}
for which the constraint surface is $\Sigma = \mathbb{R} \times \Gamma_{\can,N}$ such that $H_N$ is a smooth function on $\Gamma_{\can,N}$. Then, restricting considerations on such $\Sigma$, we can define a relational Gibbs state on the reduced system $\Gamma_{\can,N}$,

\begin{equation}
\rho_{\can} := \frac{1}{Z_{\can}} e^{-\beta H_N} \;\;,\;\; Z_{\can} = \int_{\Gamma_{\can,N}} d\mu_{\can,N} \; e^{-\beta H_N}  \; ,
\end{equation}
where $d\mu_{\can,N}$ is the Liouville measure (in local coordinates, the Lebesgue measure) on $\Gamma_{\can,N}$. This is indeed a Gibbs state which is at equilibrium with respect to the flow $X_{H_N}$ that is parametrized by the clock time $t$. Equivalently, on $\Gamma_{\phy, N}$, the state $\rho = \frac{1}{Z_{\can}} e^{-\beta (\pi_*H_N)}$ is a physical statistical state which is at equilibrium with respect to the flow generated by $X_{H_{\phy}}$ on $\Gamma_{\phy,N}$, where $\pi^* H_{\phy} = H_N$. Note that $X_{H_{\phy}} = -\pi_*(\partial_t)$ (since $\pi_*(X_C) = \pi_*(\partial_t + X_{H_N}) = 0$). \\

In the quantum systems associated with these classical systems as described in section \ref{quantise}, formal constructions of the corresponding Gibbs density operators follow through straightforwardly. The relational Gibbs state is a density operator on $\mathcal{H}_{\can,N}$ of the form $\widehat{\rho}_{\can} = \frac{1}{Z_{\can}} e^{-\beta \hat{H}_N}$. \\

It is important to notice that this equilibrium state can be obtained as the reduction of the KMS state defined in section \ref{KMS-phi} with respect to the same translation in the (now) clock variable (one of the original internal scalar field variables), after the imposition of the dynamical constraint of the theory (i.e. on-shell with respect to the fundamental dynamics of the (quantum) system), and after the deparametrizing approximations have been imposed on the same dynamical constraint.


\section{Conclusion}

In this work we have tackled the issue of defining statistical equilibrium in group field theory quantum gravity, i.e. in a complete background independent context (thus in absence of any preferred notion of time evolution) and dealing with the fundamental (candidate) microscopic degrees of freedom of a quantum spacetime (themselves not directly spatiotemporal). More specifically, we investigated the definition and construction of Gibbs states within the quantum operator formulation of group field theory for discrete gravity coupled to a number of real scalar matter fields. We have stressed how the peculiar mathematical formulation of GFT offers several advantages toward achieving our goal, also in comparison with other quantum gravity formalisms. Before discussing our explicit constructions, we have outlined the different strategies that could be followed and the different underlying principles, corresponding to different possible definitions of what is meant by statistical equilibrium. We then offered three examples. \\

The first was based on the principle of constrained maximisation of the entropy in the spirit of Jaynes' work: the constraints are a set of macroscopic observables (that one has access to), and the result of maximising the entropy under fixed values of these is to find the corresponding, least-biased distribution over the microscopic states such that the statistical averages of the given set of observables coincide with their macroscopic values. This method does not require a pre-defined transformation to define equilibrium with respect to. Although once a state is defined, one can (if one wants) extract its modular flow with respect to which it will satisfy the KMS condition (along the lines of the thermal time hypothesis). Thus it could be especially useful in quantum gravity contexts. We have observed here the general construction principles of such states, and have yet to explore their full potential. A more complete analysis and further interesting examples are left to forthcoming work \cite{kotechainprep}. As a simple illustrative example, here we have constructed a geometric volume Gibbs state characterised by a fixed average of a volume operator defined on the GFT Fock space. It was found that a direct consequence of the system being defined in such a state was the occurrence of Bose-Einstein condensation to a low-spin phase. Naturally it would be useful to consider more examples of a similar type in future works. An interesting possibility is to apply the GFT statistical mechanical framework formulated here to the case of spherical GFT condensate states in the context of quantum black hole studies like \cite{pranz1, *pranz2}, where in fact a similar construction was used in terms of the area operator, as is often considered in loop quantum gravity-inspired analysis of quantum black holes \cite{DiazPolo:2011np}. \\

The second characterisation of Gibbs states that we have considered is the KMS condition with respect to a 1-parameter group of automorphisms of the GFT algebra. After showing that in the GFT quantum system at hand with a given automorphism group, the unique KMS states are the Gibbs states with respect to the generator of the automorphism, we constructed example Gibbs states encoding equilibrium with respect to internal translations along the base manifold $G^d \times \mathbb{R}^n$. These momentum Gibbs states were structural and model-independent. Keeping in mind that the primary reason to couple scalar fields was to then use them to define a physical, relational reference frame, it was natural to seek a Gibbs state, still generated by the momentum of the scalar field but which also encoded relational dynamics defined within a deparametrized system. \\

This led us to the final example: a relational Gibbs state encoding equilibrium with respect to physical relational evolution, for a given dynamical constraint. This was based on a deparametrization procedure, and our construction was confined to a non-interacting $N$-particle GFT system, for which a single global clock was extracted. For such systems, we identified the relevant relational Gibbs density operator. In this aspect too, several things remain to be better understood. Particularly, extending the current investigation to the full GFT Fock space, thus lifting the restriction to an $N$-particle sector, is important. Also, considering examples of specific models will be valuable. Lastly, it would be interesting to investigate further the idea of a deparametrized system achieved as a certain sector of the original non-deparametrized system, at the full quantum level, via semi-classical states.


\begin{acknowledgments}
We are thankful to Goffredo Chirco, Alexander Kegeles and Seungjin Lee for many insightful discussions. We are also grateful to an anonymous referee for several helpful comments. IK is grateful to DAAD for financial support under the funding program ``Research Grants - Doctoral Programmes in Germany, 2015-16 (57129429)".
\end{acknowledgments}


\appendix

\section{Strong continuity of unitary translation groups} \label{continuityproof}

The existence of unitary groups in Fock space has been established using the invariance of the algebraic Fock state under the translation automorphisms. Here we show that the map $\underline{g} \mapsto U(\underline{g})$ is strongly continuous in $\mathcal{H}_F$, given $G^d \ni \underline{g}$. Notice that the case $G = \mathbb{R}$, $d=n$ is already included within this more general case. \\

\textbf{Claim.} $U(\underline{g})$ is a strongly continuous family of operators in $\mathcal{H}_F$, that is, $||(U(\underline{g}_1) - U(\underline{g}_2))\psi|| \to 0 \;$ as $\;\underline{g}_1 \to \underline{g}_2$, for all $\psi \in \mathcal{H}_F$, and all $\underline{g}_1, \underline{g}_2 \in G^d$.\\

\textbf{Proof.} The strategy will be to first show strong continuity at the identity $\underline{e}$, which can then be extended to all elements due to boundedness of $U$. For the first part, we begin by considering the set $\{\pi_F(W(f))\Omega_F \;| \; f \in \mathcal{H}\}$ of basis vectors of Fock space,
\begin{equation*}
||(U(\underline{g})-U(\underline{e}))\pi_F(W(f))\,\Omega_F||^2 = ||\pi_F(W(L^*_{\underline{g}} f))\Omega_F||^2 + ||\pi_F(W(f))\Omega_F||^2 - 2 \, \text{Re}\,(\pi_F(W(f))\Omega_F, \pi_F(W(L^*_{\underline{g}} f))\Omega_F) \;,
\end{equation*}
using $U(\underline{e}) = I$. Notice that, 
\begin{equation*} 
||\pi_F(W(L^*_{\underline{g}} f))\Omega_F||^2 \leq ||\pi_F(W(L^*_{\underline{g}} f))||^2\,||\Omega_F||^2 = ||\pi_F(W(L^*_{\underline{g}} f))||^2 = ||W(L^*_{\underline{g}} f)||_{\star}^2 = ||W(f)||_{\star}^2 = 1 \;,
\end{equation*}
where we have used the facts that the Fock representation is faithful in the third to last equality, and that all C*-automorphisms are norm-preserving in the penultimate equality. Also, $||.||_{\star}$ denotes the C*-norm, while $||.||$ with no subscript denotes the standard operator norm in $\mathcal{H}_F$. We thus have, $||\pi_F(W(L^*_{\underline{g}} f))\Omega_F||^2 + ||\pi_F(W(f))\Omega_F||^2 \leq 2 $, giving,
\begin{align*}
||(U({\underline{g}})-I)\pi_F(W(f))\,\Omega_F||^2 &\leq 2 - 2 \, \text{Re}\,(\pi_F(W(f))\Omega_F, \pi_F(W(L^*_{\underline{g}} f))\Omega_F)  \\
		&= 2\,\left[1 - e^{-||L^*_{\underline{g}} f-f||^2/4} \cos\left(\frac{1}{2}\text{Im}(-f,L^*_{\underline{g}} f)\right) \right]  \longrightarrow 0
\end{align*}
because $L^*_{\underline{g}} f \rightarrow f$ as ${\underline{g}} \rightarrow \underline{e}$ using continuity of the left regular representation for unimodular groups. This implies that,
\begin{equation} \label{strongcontinuitygenerator}
||(U({\underline{g}})-I)\pi_F(W(f))\,\Omega_F|| \to 0 \;\;\;\;\text{as} \;\;\;\; {\underline{g}} \to \underline{e} \;. \end{equation} 

Then, for any $\psi$ in the dense domain $D(\mathcal{H}_F) = \pi_F(\mathcal{A})\Omega_F$, written as a general linear superposition $\psi = \pi_F(A)\Omega_F = \pi_F(\sum_i c_i W(f_i)) \Omega_F = \sum_i c_i \, \pi_F(W(f_i)) \Omega_F$, we have,
\begin{align*}
||(U({\underline{g}})-I)\psi|| &= ||(U({\underline{g}})-I)\sum_i c_i \, \pi_F(W(f_i)) \Omega_F||  \\
			&\leq \sum_i |c_i| \, ||(U({\underline{g}})-I) \, \pi_F(W(f_i)) \Omega_F||  \longrightarrow 0
\end{align*}
as ${\underline{g}} \to \underline{e}$, using result $\eqref{strongcontinuitygenerator}$ for each $i$.  Thus, $U({\underline{g}})$ is continuous in the strong operator topology in $D(\mathcal{H}_F)$. Since it is bounded, it can be extended to the whole $\mathcal{H}_F$.
$\hfill \square$  


\section{Mathematical checks for volume Gibbs operator} \label{volumechecks}

For simplicity let us begin with first checking that the state $e^{-\beta V}$ is positive and trace-class. Technically, this is like the canonical ensemble and is accompanied by the constraint $N_{\text{tot}} = \sum_{\underline{\chi}} n_{\underline{\chi}}$. Since the total number of particles is constrained to be $N_{\text{tot}}$, the relevant Hilbert space here is $\mathcal{H}^{\otimes N_{\text{tot}}}$, seen as a restriction of $\mathcal{H}_F$ to the $N_{\text{tot}}$-particle sector. The following computations use the orthonormal occupation number basis $\{\ket{\{n_{\underline{\chi}_i}\}}\}$ of $\mathcal{H}_F$. For convenience, denote $\underline{\chi}_i \equiv i$ here. \\


\textbf{Claim 1.1.} Operator $e^{-\beta V}$, for $0 < \beta < \infty$ and $V$ as defined in \eqref{vol}, is bounded in the operator norm on $\mathcal{H}_F$. \\

\textbf{Proof 1.1.} An operator $A$ is called bounded when there exists a real $k \geq 0 $ such that $||A\psi|| \leq k||\psi|| $ for all $\psi$ in the relevant Hilbert space. Considering first the basis vectors,
\begin{equation} \nonumber
||e^{-\beta V} \ket{\{n_i\}}|| = ||e^{-\beta \sum_i v_i n_i} \ket{\{n_i\}}|| =  e^{-\beta \sum_i v_i n_i} \; || \ket{\{n_i\}}|| \;\;.
\end{equation}

Then, for a generic state $\ket{\psi} = \sum_{\{n_i\}} \langle \{n_i\} | \psi \rangle \ket{\{n_i\}} \equiv  \sum_{\{n_i\}} c_{\{n_i\}} \ket{\{n_i\}} $, we have,
\begin{align*} 
||e^{-\beta V} \psi||^2 &= || \sum_{\{n_i\}} c_{\{n_i\}} e^{-\beta \sum_i v_i n_i} \ket{\{n_i\}} ||^2 \;= \sum_{\{n_i\}, \{n_j\}}  \bar{c}_{\{n_j\}} c_{\{n_i\}} e^{-\beta \sum_j v_j n_j} e^{-\beta \sum_i v_i n_i} \langle \{n_j\} | \{n_i\} \rangle \\ 
&= \sum_{\{n_i\}} | c_{\{n_i\}} |^2  e^{- 2 \beta \sum_i v_i n_i} \leq \sum_{\{n_i\}} | c_{\{n_i\}} |^2 = || \psi ||^2 \;\;,
\end{align*}


using orthonormality of basis, and $\; \beta \sum_i v_i n_i \geq 0 \; \Rightarrow \; 0 < e^{- 2 \beta \sum_i v_i n_i} \leq 1 \;. \hfill \square$ \\


\textbf{Claim 1.2.} The bounded operator $e^{-\beta V}$ is positive on $\mathcal{H}_F$. \\

\textbf{Proof 1.2.} A bounded operator $A$ is positive if $\bra{\psi} A \ket{\psi} \geq 0$ for all $\psi$ in the relevant Hilbert space. For the basis vectors we straightforwardly have,
\begin{equation}\nonumber
\bra{\{n_i\}} e^{-\beta V} \ket{\{n_i\}} = e^{-\beta \sum_i v_i n_i} ||\ket{\{n_i\}}||^2 \geq 0 \;.
\end{equation}
Then, for any state $\psi \in \mathcal{H}^{\otimes N_{\text{tot}}}$,
\begin{equation} \nonumber
\bra{\psi} e^{-\beta V} \ket{\psi} = \sum_{\{n_i\}, \{n_j\}}  \bar{c}_{\{n_j\}} c_{\{n_i\}} \bra{\{n_i\}} e^{-\beta \sum_j v_j n_j} \ket{\{n_j\}} = \sum_{\{n_i\}}  |c_{\{n_i\}}|^2 e^{-\beta \sum_i v_i n_i} \geq 0 \;.
\end{equation}
$\hfill \square$ 


\textbf{Claim 1.3.} Operator $e^{-\beta V}$ is trace-class on $\mathcal{H}_F$. \\

\textbf{Proof 1.3.} The trace is,
\begin{equation*}
\Tr(e^{-\beta V}) \;=\; \sum_{\{n_{\underline{\chi}}\}} e^{-\beta \sum_{\underline{\chi}} v_{\underline{\chi}}n_{\underline{\chi}} } \; \equiv \;\sum_{\{n_{\underline{\chi}}\}} e^{-\beta \; \mathcal{V}_{\{n_{\underline{\chi}}\}} }
\end{equation*}

where the sum is over all possible ways of arranging $N_{\text{tot}}$ particles into an arbitrary number of boxes labelled by $\underline{\chi}$. Now, the configuration with the lowest total volume will be the one in which all $N_{\text{tot}}$ particles occupy the single-particle ground state with volume $v_0$. This is the ground state $\ket{N_{\text{tot}},0,0,...}$ of the total volume operator $V$, with eigenvalue $\mathcal{V}_0 = N_{\text{tot}}v_0$. The highest contribution to the above sum comes from this term. We can now separate this contribution to rewrite the series as,
\begin{equation*}
\Tr(e^{-\beta V})\;\;=\;\; e^{-\beta \mathcal{V}_0} \;\;+ \sum_{\ket{\{n_{\underline{\chi}}\}} \neq \ket{N_{\text{tot}},0,...}} e^{-\beta \mathcal{V}_{\{n_{\underline{\chi}}\}} } 
\end{equation*}
where now all $e^{-\beta \mathcal{V}_{\{n_{\underline{\chi}}\}} } < e^{-\beta \mathcal{V}_0}$. Now, we rearrange the states in the sum in increasing values of total volume eigenvalues (and denote these with tildes), so that 
\begin{equation*}
\sum_{\ket{\{n_{\underline{\chi}}\}} \neq \ket{N_{\text{tot}},0,...}} e^{-\beta \mathcal{V}_{\{n_{\underline{\chi}}\}} } \;\; = \;\; \sum_{\widetilde{\mathcal{V}}_{\{n_{\underline{\chi}}\}_l} > \mathcal{V}_0} e^{-\beta \widetilde{\mathcal{V}}_{\{n_{\underline{\chi}}\}_l} } \;\;\equiv \;\; \sum_{\widetilde{\mathcal{V}}_l > \mathcal{V}_0} e^{-\beta \widetilde{\mathcal{V}}_l } \;\;\;.
\end{equation*}

Here, $l \in \{1, 2, 3, ... \}$ labels the reorganised list of multi-particle states in ascending order of their volume eigenvalues, $\widetilde{\mathcal{V}}_l \leq \widetilde{\mathcal{V}}_{l+1}$, where equality denotes degeneracy of adjacent states. We have thus rearranged the series such that each exponential term is less than or equal to the previous one. This series converges (by ratio test), \begin{equation}\nonumber
r \equiv \lim_{l \rightarrow \infty} \frac{e^{-\beta \widetilde{\mathcal{V}}_{l+1} }}{e^{-\beta \widetilde{\mathcal{V}}_l }} = \lim_{l \rightarrow \infty} e^{-\beta (\widetilde{\mathcal{V}}_{l+1} - \widetilde{\mathcal{V}}_l)} < 1 \;\;.
\end{equation}
$\hfill \square$ 


The same properties are now verified for the operator of interest, $e^{-\beta(V-\mu N)}$, for positive $\beta$ as above. The relevant Hilbert space is now the full $\mathcal{H}_F$ as the particle number is allowed to fluctuate. \\


\textbf{Claim 2.1.} Operator $e^{-\beta(V-\mu N)}$, for $0 < \beta < \infty$ and $\mu \leq v_0$, is bounded in the operator norm on $\mathcal{H}_F$. \\

\textbf{Proof 2.1.} For a generic state $\psi \in \mathcal{H}_F$,
 \begin{equation*} 
||e^{-\beta (V - \mu N)} \psi||^2 = \sum_{\{n_i\}} | c_{\{n_i\}} |^2  e^{- 2 \beta \sum_i (v_i - \mu) n_i} \leq \sum_{\{n_i\}} | c_{\{n_i\}} |^2 = ||\psi||^2
\end{equation*}
since, $\mu \leq v_0 \Rightarrow \mu \leq v_i$ for all $i$ $\Rightarrow 0 < e^{- 2 \beta \sum_i (v_i - \mu) n_i} \leq 1$.$\hfill \square$ 


\textbf{Claim 2.2.} The bounded operator $e^{-\beta(V-\mu N)}$ is positive on $\mathcal{H}_F$. \\

\textbf{Proof 2.2.} For a generic state $\psi$ in the Fock space, we have,
\begin{equation*} 
\bra{\psi} e^{-\beta (V - \mu N)} \ket{\psi} = \sum_{\{n_i\}}  |c_{\{n_i\}}|^2 e^{-\beta \sum_i (v_i - \mu) n_i} \geq 0 \;.
\end{equation*} $\hfill \square$


\textbf{Claim 2.3.} Operator $e^{-\beta (V-\mu N)}$ is trace-class on $\mathcal{H}_F$. \\

\textbf{Proof 2.3.} All arguments as made above in Proof 1.3 will apply here, with $v_{\underline{\chi}}$ replaced everywhere by $(v_{\underline{\chi}} - \mu)$, for any $\mu \leq v_0$. $\hfill \square$


\section{Strong continuity of map $U_X$} \label{cont}

\textbf{Claim.} Given a continuous map $g_X: \mathbb{R} \to \mathfrak{G} : t \to g_X(t)$ and a strongly continuous map $U: \mathfrak{G} \to \mathcal{U}(\mathfrak{H}) : g \mapsto U(g)$, then the map $U_X := U \circ g_X : t \mapsto U_X(t)$ is strongly continuous. \\

\textbf{Proof.} Strong continuity of $U$ means, $||(U(g_1) - U(g_2))\psi|| \to 0$ as $g_1 \to g_2$, for any $g_1,g_2 \in \mathfrak{G}$ and all $\psi \in \mathfrak{H}$. Then, for any $t_1,t_2 \in \mathbb{R}$ and all $\psi \in \mathfrak{H}$, we have
\begin{align*}
||(U_X(t_1) - U_X(t_2))\psi|| &= ||(U(g_X(t_1)) - U(g_X(t_2)))\psi|| \\
&= ||(U(g_1) - U(g_2))\psi||
\end{align*}
where $g_1 \equiv g_X(t_1)$ and $g_1 \equiv g_X(t_2)$ are arbitrary elements on the curve $g_X(t) \in \mathfrak{G}$. Continuity of map $g_X$ means, $t_1 \to t_2$ implies $g_1 \to g_2$. Then, using strong continuity of $U$, we have as $t_1 \to t_2$,
\begin{equation*}
||(U(g_1) - U(g_2))\psi|| \to 0  \;. 
\end{equation*} $\hfill \square$


\bibliographystyle{unsrt}
\bibliography{ref1}

\end{document}